\documentclass[journal,twocolumn]{IEEEtran}
\usepackage{amsfonts}
\usepackage[tbtags]{amsmath}
\usepackage{graphicx,subfigure}
\usepackage{color}      
\usepackage{epsfig}
\usepackage{psfig}
\usepackage{amssymb}
\usepackage{tipa}
\usepackage{bbm}

\def\endproof{\hspace*{\fill}~$\blacksquare$}
\long\def\comment#1{}

\newcommand{\beq}{\begin{equation}}
\newcommand{\eeq}{\end{equation}}
\newcommand{\beqno}{\begin{equation*}}
\newcommand{\eeqno}{\end{equation*}}

\newcommand{\bes}{\begin{split}}
\newcommand{\ees}{\end{split}}
\newcommand{\bdm}{\begin{displaymath}}
\newcommand{\edm}{\end{displaymath}}

\newcommand{\goes}{\rightarrow}

\newtheorem{theorem}{Theorem}

\newtheorem{lemma}{Lemma}

\newtheorem{definition}{Definition}

\newcommand{\bd}{\begin{definition}}
\newcommand{\ed}{\end{definition}}

\newcommand{\bv}{\begin{vugraph}}
\newcommand{\ev}{\end{vugraph}}
\newcommand{\bi}{\begin{itemize}}
\newcommand{\ei}{\end{itemize}}
\newcommand{\ben}{\begin{enumerate}}
\newcommand{\een}{\end{enumerate}}

\newcommand{\bean}{\begin{eqnarray*} }
\newcommand{\eean}{\end{eqnarray*} }
\newcommand{\bea}{\begin{eqnarray} }
\newcommand{\eea}{\end{eqnarray} }

\newcommand{\ba}{\begin{array} }
\newcommand{\ea}{\end{array} }

\def\calR{\mathbb{R}}
\begin{document}

\title{Minimum cost distributed source coding over a network}

\author{Aditya Ramamoorthy \thanks{Aditya Ramamoorthy is with the Department of Electrical and Computer Engineering, Iowa State University, Ames IA 50011, USA (email: adityar@iastate.edu). The material in this paper was
presented in part at the IEEE International Symposium on
Information Theory, Nice, France June 2007}}
\maketitle
\begin{abstract}
This work considers the problem of transmitting multiple
compressible sources over a network at minimum cost. The aim is to
find the optimal rates at which the sources should be compressed
and the network flows using which they should be transmitted so
that the cost of the transmission is minimal. We consider networks
with capacity constraints and linear cost functions. The problem
is complicated by the fact that the description of the feasible
rate region of distributed source coding problems typically has a
number of constraints that is exponential in the number of sources.
This renders general purpose solvers inefficient. We present a
framework in which these problems can be solved efficiently by
exploiting the structure of the feasible rate regions coupled with
dual decomposition and optimization techniques such as the
subgradient method and the proximal bundle method.
\end{abstract}
\IEEEpeerreviewmaketitle
\section{Introduction}
\label{intro-sec} In recent years the emergence of sensor networks
\cite{pottieK00} as a new paradigm has introduced a number of
issues that did not exist earlier. Sensor networks have been
considered among other things by the military for battlefields, by
ecologists for habitat monitoring and even for extreme event
warning systems. These networks consist of tiny, low-power nodes
that are typically energy constrained. In general, they also have
low-computing power. Thus, designing efficient sensor networks
requires us to address engineering challenges that are
significantly different from the ones encountered in networks such
as the Internet. One unique characteristic of sensor networks is
that the data that is sensed by different sensor nodes and relayed
to a terminal is typically highly correlated. As an example
consider a sensor network deployed to monitor the temperature or
humidity levels in a forest. The temperature is not expected to
vary significantly over a small area. Therefore we do expect that
the readings corresponding to nearby sensors are quite correlated.
It is well-known that the energy consumed in transmission by a
sensor is quite substantial and therefore efficient low power
methods to transfer the data across the network are of interest.
This leads us to investigate efficient techniques for exploiting
the correlation of the data while transmitting it across the
network. There are multiple ways in which the correlation can be
exploited.
\begin{itemize}
\item[a)] The sensor nodes can communicate amongst themselves to
inform each other of the similarity of their data and then
transmit only as much data as is required. This comes at the cost
of the overhead of inter-sensor communication. \item[b)] The
sensors can choose to act independently and still attempt to
transmit the compressed data. This strategy is likely to be more
complicated from the receiver's point of view.
\end{itemize}
Usually the terminal to which the data is transmitted has
significantly more resources (energy, computing power etc.). Thus,
the latter solution is more attractive from a network resource
efficiency point of view. The question of whether the distributed
compression of correlated sources can be as efficient as their
compression when the sources communicate with each other was first
considered and answered in the affirmative by Slepian and Wolf in
their famous paper \cite{slepianwolf}. A number of authors
\cite{pradhandiscus}\cite{stankovicLXG06} have investigated the
construction of coding techniques that achieve the Slepian-Wolf
bounds and also proposed their usage in sensor networks
\cite{xiongspmag}.

New paradigms have also emerged recently in the area of network
information transfer. Traditionally information transfer over
networks has been considered via routing. Data packets from a
source node are allowed to be replicated and forwarded by the
intermediate nodes so that terminal nodes can satisfy their demands.
However, network coding offers an interesting alternative where
intermediate nodes in a network have the ability to forward
functions of incoming packets rather than copies of the packets.
The seminal work of Ahlswede et al. \cite{al} showed that network
coding achieves the capacity of single-source multiple-terminal
multicast where all the terminals are interested in receiving the
same set of messages from the source. This was followed by a
number of works that presented constructions and bounds for
multicast network codes \cite{lc}\cite{jaggiSCEEJT05}. More
recently, there has been work \cite{songY03}\cite{harveyKL06} on
characterizing rate regions for arbitrary network connections
where the demands of the terminals can be arbitrary.

Given these developments in two different fields, a natural
question to ask is how can one transmit compressible sources over
a network using network coding and whether this can be done
efficiently. This problem was
considered by Song and Yeung \cite{songY01} and Ho et al. \cite{tracey_ciss}. 
They showed that as long as the minimum cuts between all nonempty
subsets of sources and a particular terminal were sufficiently
large, random linear network coding over the network followed by
appropriate decoding at the terminals achieves the Slepian-Wolf
bounds. The work of Ramamoorthy et al. \cite{adisepDSC}
investigated the performance of separate source and network codes
and showed that separation does not hold in general. 
Both these papers only considered capacity constraints on the
edges of the network and did not impose any cost associated with
edge usage.
\par In the networking context the problem of minimum cost network
flow has been widely investigated. Here, every edge in the network
has a cost per unit flow associated with it. The cost of a given
routing solution is the sum of the costs incurred over all the
links. The problem is one of finding network flows such that the
demand of the terminals is satisfied at minimum cost. This problem
has been very well investigated in the routing context \cite{rk}.
The problem of minimum cost multicast using network coding was
considered by Lun et al. \cite{lunmincost} and they presented
centralized and distributed solutions to it.

In this paper we consider the problem of minimum cost joint rate
and flow allocation over a network that is utilized for
communicating compressible sources. We consider the scenario when
the compression is to be performed in a distributed manner. The
sources are not allowed to communicate with each other. The main issue with joint rate and flow allocation
is that typically the feasible rate region for the recovery of the
sources (e.g. the Slepian-Wolf region) is described by a number of
inequalities that is exponential in the number of sources. Thus, using a regular LP solver for solving the corresponding
linear programming problem will be inefficient. In our work, we only consider networks where the links are independent and where transmission up to the link's capacity is assumed to be error free. In general, the capacity region characterization of more complex networks such as wireless networks will need to take account issues such as interference. Moreover, it would introduce related issues such as scheduling. We do not consider these problems in this work.
\subsection{Main Contributions}
The main contributions of this paper are as follows. We present a
framework in which minimum cost problems that involve transmitting
compressible sources over a network in a distributed manner can be
solved efficiently. We consider general linear cost functions,
allow capacity constraints on the edges of the network and
consider the usage of network coding. The following problems are considered.
\begin{itemize}
\item[a)] {\it Slepian-Wolf over a network.}  The sources are
assumed to be discrete and memoryless and they need to be
recovered losslessly \cite{slepianwolf} at the terminals of the
network. We address the problem of jointly finding the operating
rate vectors for the sources and the corresponding network flows
that allow lossless recovery at the terminals at minimum cost.
\item[b)] {\it Quadratic Gaussian CEO over a network.} A Gaussian
source is observed by many noisy sensors and needs to be recovered
at the terminal subject to a quadratic distortion constraint
\cite{viswanathanB97}. We present a solution to the problem of
joint rate and network flow allocation that allows recovery at the
terminal at minimum cost.\item[c)] {\it Lifetime maximization of
sensor networks with distortion constraints.} A Gaussian source
observed by many noisy sensors needs to be recovered at the
terminal with a certain fidelity. We are interested in finding
routing flows that would maximize the lifetime of the network.
\end{itemize}
We demonstrate that these problems can be solved efficiently by
exploiting the structure of the feasible rate regions coupled with
dual decomposition techniques and subgradient methods
\cite{nonlinear_bert}.

\subsection{Related Work}
\label{sec:related_work}
Problems of a similar flavor have been examined in several papers. Cristescu et al. considered the Networked Slepian-Wolf problem \cite{cristescuLV05} and the case of lossy correlated data gathering over a network \cite{cristescuL06}, but did not impose capacity constraints on the edges. Their solutions only considered very specific types of cost functions. The work of Li \& Ramamoorthy \cite{LiR08,LiR08jnl} and Roumy \& Gesbert \cite{roumyG07} considered a rate allocation under pairwise constraints on the distributed source code used for compression. The work of Liu et al. \cite{liuATZ06} and \cite{liuATZ_jnl} considers a related problem, where they seek to minimize the total communication cost of a wireless sensor network with a single sink. They show that when the link communication costs are convex, then the usage of Slepian-Wolf coding and commodity flow routing is optimal. Moreover, they introduce the notion of distance entropy, and show that under certain situations the distance entropy is the minimum cost achieved by Slepian-Wolf coding and shortest path routing. They also propose hierarchical transmission schemes that exploit correlation among neighboring sensor nodes, and do not require global knowledge of the correlation structure. These schemes are shown to be order-optimal in specific situations. The main difference between our work and theirs, is the fact that we consider network coding and networks with multiple terminals. Moreover, in the case of general convex link cost functions, their focus is on showing that Slepian-Wolf coding and commodity flow routing is optimal. They do not consider the problem of actually finding the optimal flows and rates.

A problem formulation similar to ours was introduced by Barros et al. \cite{barrosservetto} but they did not present an efficient solution to it. The problem of exponentially many constraints has been noted by other authors as well \cite{kansalramamoorthy}\cite{LiD06}.

The approach in our work is inspired by
the work of Yu et al. \cite{yuyuan}. However since our cost
functions only penalize the usage of links in the network, we are
effectively able to exploit the structure of the feasible rate
region to make our overall solution efficient. In addition we
explicitly derive the dual function and the corresponding update
equations for maximizing it based on the specific structure of the
rate region. Furthermore, we consider applications in network
coding and lifetime maximization in sensor networks that have not
been considered previously. In concurrent and independent work
\cite{cuiHC07} presented some approaches similar to ours (see also \cite{leeMHGR07}, where the case of two sources is discussed). However
our approach has been applied to the minimum cost quadratic
Gaussian CEO problem over a network and lifetime maximization with
distortion constraints that were not considered in \cite{cuiHC07}.

A reviewer has pointed out that the problem of generalizing the Slepian-Wolf theorem to the network case was first considered by Han \cite{han80} in 1980. However, in \cite{han80} only networks with a single terminal were considered. In the single terminal case the corresponding flows can be supported by pure routing. Interestingly, in the same paper, Han references the work of Fujishige \cite{fujishige78} that studies the optimal independent flow problem (this was also pointed by the same reviewer). Fujishige's work considers a network flow problem that has polymatroidal \cite{tsehanly_1} constraints for the source values and the terminal values. In particular, if there is only one terminal, then this algorithm provides an efficient solution to the minimum cost Slepian-Wolf problem over a network. However, it is unclear whether it can be extended to the case of multiple terminals and network coding. We discuss Fujishige's work in more detail in Section \ref{sec:sw-over-network}, after the precise problem has been formulated.

\par This paper is organized as follows. 
Section \ref{prob-form-sec} overviews the notation and the broad setup under consideration in this paper.
Section \ref{sec:sw-over-network} formulates and solves the
minimum cost Slepian-Wolf problem over a network, Section
\ref{sec:ceo_over_network} discusses the quadratic Gaussian CEO
problem over a network and Section \ref{sec:lifetime_over_network}
presents and solves the problem of lifetime maximization of sensor
networks when the source needs to be recovered under distortion
constraints. Each of these sections also present simulation
results that demonstrate the effectiveness of our method. Section \ref{sec:conclusions} discusses the conclusions and future work.

\section{Preliminaries}
\label{prob-form-sec} 

In this section we introduce the basic problem setup that shall be
used in the rest of this paper. In subsequent sections we shall be
presenting efficient algorithms for solving three different
problems that fall under the umbrella of distributed source coding
problems over a network. We shall present the exact formulation of
the specific problem within those sections. We are given the
following.

\begin{itemize}
\item[a)] A directed acyclic graph $G = (V, E, C)$ that represents
the network. Here $V$ represents the set of vertices, $E$ the set
of edges and $C_{ij}, (i,j) \in E$ is the capacity of the edge
$(i,j)$ in bits/transmission. The edges are assumed to be
error-free and the capacity of the edges is assumed to be
rational. We are also given a set of source nodes $S \subseteq V$
where $|S| = N_S$ and a set of terminal nodes $T \subseteq V$
where $|T| = N_R$. Without loss of generality we assume that the
vertices are numbered so that the vertices $1, 2, \dots, N_S$
correspond to the source nodes. \item[b)] A set of sources $X_1,
X_2, \dots X_{N_S}$, such that the $i^{th}$ source is observed at
source node $i \in S$. The values of the sources are drawn from
some joint distribution and can be either continuous or discrete.
\end{itemize}

Based on these we can define the capacity region of the terminal
$T_j \in T$ with respect to $S$ as

\begin{equation*}
C_{T_j} = \big{\{}(R_1,\ldots,R_{N_S}) :
\forall B \subseteq S,\sum_{i \in B} R_i \leq \textrm{min-cut}(B,
T_j)\big{\}}.
\end{equation*}
Thus, $C_{T_j}$ consists of inequalities that define the maximum
flow (or minimum cut) from each subset of $S$ to the terminal
$T_j$. A rate vector $(R_1,\ldots,R_{N_S}) \in C_{T_j}$ can be
transmitted from the source nodes $1, \dots , N_S$ to terminal
$T_j$ via routing \cite{rk}. In the subsequent sections we shall
consider different minimum cost problems involving the
transmission of the sources over the network to the terminals.


\section{Minimum cost Slepian-Wolf over a network}
\label{sec:sw-over-network} Under this model, the sources are
discrete and memoryless and their values are drawn i.i.d. from a
joint distribution $p(x_1,\dots,x_{N_S})$. 
The $i^{th}$ source node only observes $X_i$ for $i \in S$. The
different source nodes operate independently and are not allowed
to communicate. The source nodes want to transmit enough
information using the network to the terminals so that they can recover the original sources, losslessly.

This problem was first investigated in the seminal paper of
Slepian and Wolf \cite{slepianwolf} where they considered the
sources to be connected to the terminal by a direct link and the
links did not have capacity constraints. The celebrated result of
\cite{slepianwolf} states that the independent source coding of
the sources $X_i, i = 1, \dots, N_S$ can be as efficient as joint
coding when the sources need to be recovered error-free at the
terminal.
\par Suppose that for the classical Slepian-Wolf problem, the rate of the $i^{th}$ source is $R_i$.
Let $X_{B}$ denote the vector of sources $(X_{i_1}, X_{i_2},
\dots, X_{i_{|B|}})$, for $i_k \in B, k = 1, \dots, |B|$. The
feasible rate region for this problem is given by
\begin{equation*}
\mathcal{R_{SW}} = \{(R_1, \dots,R_{N_S}):
\forall B \subseteq S, \sum_{i \in B} R_i \geq H(X_{B}|X_{B^c})\}
\end{equation*}
The work of Csisz\'{a}r \cite{csislinear} showed that linear codes are
sufficient to approach the Slepian-Wolf (henceforth S-W) bounds
arbitrarily closely.
\par Note that the original S-W problem does not consider
the sources to be communicating with the terminal (or more
generally multiple terminals) over a network. Furthermore, there
are no capacity constraints on the edges connecting the sources
and the terminal. In situations such as sensor networks, where the
sensor nodes are typically energy-constrained, we would expect the
source nodes to be in communication with the terminal node over a
network that is both capacity and cost-limited. Capacity
constraints may be relatively strict since a significant amount of
power is consumed in transmissions. The costs of using different
links could be used to ensure that a certain part of the network
is not overused resulting in non-uniform depletion of resources.
Thus the problem of transmitting correlated data over a network
with capacity constraints at minimum cost is of interest. We
define an instance of the S-W problem over a network by
$P = <\mathcal{R_{SW}}, G, S, T>$.


The transmission schemes based on linear codes (such as those in
\cite{csislinear}) are based on block-wise coding, i.e., each source
encodes $N$ source symbols at a time. An edge with capacity
$C_{ij}$ bits/transmission can transmit $\lfloor NC_{ij} \rfloor$
bits per block. Conceptually, each edge can be regarded as
multiple unit capacity edges, with each unit capacity edge capable
of transmitting one bit per block. When communicating a block of
length $N$, we consider the graph $G^N=(V,E,\lfloor NC\rfloor)$,
or equivalently the graph $(V,E_N,\mathbbm{1})$ (where
$\mathbbm{1}$ denotes a vector of ones) where $E_N$ splits each
edge from $E$ into unit capacity edges.

\par To facilitate the problem formulation we construct an augmented graph
$G^*$ where we append a virtual super source node $s^{*}$ to $G$, so that
\begin{eqnarray*}
V^{*} &=& V \cup \{s^{*}\},\\
E^{*} &=& \{(s^{*}, v) |~ v \in S\}
\cup E, \text{~and}\\
C^*_{ij} &=& \begin{cases} C_{ij} & (i,j) \in E,\\ H(X_j)  &
\text{~if~} i = s^* \text{~and~} j \in S. \end{cases}
\end{eqnarray*}

We let $G^{*} = (V^{*},E^{*}, C^{*})$.

\begin{definition}
\label{def:feas_SW} {\em Feasibility.} Consider an instance of the
S-W problem over a network, $P = <\mathcal{R_{SW}}, G, S, T>$. Let
$C_{T_i}$ be the capacity region of each receiver $T_i \in T$ with
respect to $S$. If
\begin{equation*}
\mathcal{R_{SW}} \cap C_{T_i} \neq \emptyset, \forall i = 1,
\ldots, N_R,
\end{equation*}
then the feasibility condition is said to be satisfied and $P$ is
said to be feasible.
\end{definition}
The next theorem (from \cite{tracey_ciss}) implies that as long as the feasibility condition is satisfied, random linear network coding over $G^{N}$ followed
by appropriate decoding at $T_i$ suffices to reconstruct the
sources $X_1, X_2, \dots, X_{N_S}$ error-free at $T_i$.
\begin{theorem}
{\em Sufficiency of the feasibility condition \label{thm:
tracey_ciss_thm} \cite{tracey_ciss}}. Consider an instance of the
S-W problem over a network, $P = <\mathcal{R_{SW}}, G, S, T>$. If
the feasibility condition (Definition \ref{def:feas_SW}) is
satisfied, then random linear network coding over $G^{N}$ followed
by minimum-entropy \cite{csislinear} or maximum-likelihood
decoding can recover the sources at each terminal in $T$ with the
probability of decoding error going to $0$ as $N \rightarrow
\infty$.
\end{theorem}
The proof of the necessity of the feasibility condition can be
found in \cite{han80}.


It follows that if $C_{T_i} \cap \mathcal{R_{SW}} \neq \phi$ for
all $T_i \in T$, it is sufficient to perform random linear network
coding over a subgraph of $G$ where the feasibility condition continues to be satisfied. The question then becomes, how do we choose appropriate subgraphs?

For this purpose, we now present the formulation of the minimum cost S-W problem over a network.
%

\par 
Let $x_{ij}^{(T_k)}$ represent the flow variable for edge $(i,j)$
over $G^*$ corresponding to the terminal $T_{k}$ for $T_k \in T$
and $z_{ij}$ represent the max-of-flows variable, $\max_{T_k \in T} x_{ij}^{(T_k)}$ for edge $(i,j)$. Note that under network coding the physical flow on edge $(i,j)$ will be $z_{ij}$. The variable $x_{ij}^{(T_k)}$, represents the virtual flow variable over edge $(i,j)$ for terminal $T_k$ \cite{lunmincost}.

We introduce variables $R_i^{(T_k)},i = 1, \dots, N_S $ that
represent the operating S-W rate variables for each terminal. Thus
$R^{(T_k)} = (R_{1}^{(T_k)}, R_{2}^{(T_k)}, \dots,
R_{N_S}^{(T_k)})$ represents the rate vector for terminal $T_k$. Let $f_{ij} > 0, (i,j) \in E$, $f_{ij} = 0, (i,j) \in E^* \setminus E$ represent the cost for transmitting at a unit flow over edge $(i,j)$.
We are interested in the following optimization problem that we
call {\it MIN-COST-SW-NETWORK}.

\begin{equation}
\begin{split}
&\text{minimize~} \sum_{(i,j) \in E} f_{ij}z_{ij} \\
&\text{s. t.~~} 0 \leq x_{ij}^{(T_k)} \leq z_{ij} \leq C_{ij}^* ,
\text{~~} (i,j) \in E^{*}, T_k \in T
\label{eq:min-cost-sw-upper-bd} \raisetag{10pt}
\end{split}
\end{equation}
\begin{align}
\label{eq:min-cost-sw-flow-balance} &\sum_{\{j|(i,j) \in
E^{*}\}} x_{ij}^{(T_k)} - \sum_{\{j | (j,i) \in E^{*}\}}
x_{ji}^{(T_k)} =
\sigma_i^{(T_k)}, \\
& \text{for} ~ i \in V^{*}, T_k \in T, \nonumber
\\ \label{eq:min-cost-sw-lower-bd} &x_{s^* i}^{(T_k)} \geq R_i^{(T_k)}, \text{~~for~} i \in
S, T_k \in T \\
\label{eq:min-cost-sw-region} &R^{(T_k)} \in \mathcal{R_{SW}},
\text{~~for~} T_k \in T
\end{align}
where
\begin{equation}
\label{eq:min-cost-sw-sum-rate} \sigma_i^{(T_k)} = \begin{cases}
H(X_1, X_2,\dots, X_{N_S}) &
\text{~~if~} i = s^*\\
-H(X_1, X_2,\dots, X_{N_S}) & \text{~~if~} i = T_k \\
0 & \text{~~otherwise}
\end{cases}
\end{equation}
%

The constraints in (\ref{eq:min-cost-sw-upper-bd}),
(\ref{eq:min-cost-sw-flow-balance}) and
(\ref{eq:min-cost-sw-sum-rate}) are precisely the formulation of
the minimum cost single-source multiple terminal multicast with
network coding for a total rate of $H(X_1, X_2, \dots, X_{N_S})$.
The constraint (\ref{eq:min-cost-sw-lower-bd}) enforces the flow
$x_{s^* i}^{(T_k)}$ (corresponding to terminal $T_k$ from $s^*$
through source $i$) to be at least $R_i^{(T_k)}$. Constraint (\ref{eq:min-cost-sw-region}) ensures that the rate
vectors $R^{(T_k)}$ belong to the Slepian-Wolf region
$\mathcal{R_{SW}}$. A proof that the total rate can be fixed to be
exactly $H(X_1, X_2, \dots, X_{N_S})$ for each terminal can be
found in Appendix-I.



Suppose there exists a feasible solution $z$, $(x^{(T_k)},
R^{(T_k)})$ for $T_k \in T$ to {\it MIN-COST-SW-NETWORK}. Let
\begin{eqnarray*}
V^{*z} &=& V, \\
E^{*z} &=& \{(i,j) \in E^{*} |~ z_{ij}
> 0 \}, \text{~and} \\
C^{*z}_{ij} &=& \begin{cases} z_{ij} &\text{~if~} (i,j) \in E^{*z}\\
0& \text{~otherwise}. \end{cases}
\end{eqnarray*}


We define the subgraph of $G^{*}$ induced by $z$ 
to be the graph $G^{*}_z = (V^{*z}, E^{*z}, C^{*z})$ and the
corresponding graph over block length $N$ as $G^{*N}_z = (V^{*z},
E^{*z}_N, \mathbbm{1})$. The subgraphs induced by $x^{(T_k)}$ can
be defined analogously. We now show that if {\it
MIN-COST-SW-NETWORK} is feasible then the subgraph induced by the
feasible $z$ continues to satisfy the condition in definition
\ref{def:feas_SW} and therefore it suffices to perform random
linear network coding over this subgraph followed by appropriate
decoding at the terminals to recover the sources.

\begin{lemma}
\label{feas-lemma} Suppose that there exists a feasible solution
$z$, $(x^{(T_k)}, R^{(T_k)})$ for $T_k \in T$
to {\it MIN-COST-SW-NETWORK}. Then, random linear network coding
over the subgraph $G^{*N}_z$ induced by $z$ followed by maximum
likelihood decoding at the terminals can recover the sources $X_i,
i \in S$ at each terminal in $T$ as $N \rightarrow \infty$.
\end{lemma}

\emph{Proof:} To simplify the presentation we assume that all
$C_{ij}, (i,j) \in E^*$ and $H(X_B/X_{B^{c}}), B \subseteq S$ are
rational and the block length $N$ is large enough so that $N
C_{ij}$ and $ N H(X_B/X_{B^{c}})$ are integral. For each terminal
$T_k$ we shall show that $\textrm{min-cut} (B, T_k) \geq N H(X_B |
X_{B^{c}})$ over $G^{*N}_z$ and then use Theorem \ref{thm:
tracey_ciss_thm}.
\par Consider a terminal $T_1 \in T $. We are given the existence of a feasible solution
$(z, x, R)$ from which we can find the corresponding flow for
$T_1$ denoted by $x^{(T_1)}$. Now consider the subgraph of
$G^{*N}$ induced by $x^{(T_1)}$. Since $x^{(T_1)}$ is feasible, it
supports a rate of $N H(X_1, X_2,\dots, X_{N_S})$ from $s^*$ to
$T_1$ which implies (using Menger's theorem \cite{vanlintW}) that
there exist $N H(X_1, X_2,\dots, X_{N_S})$ edge-disjoint paths
from $s^*$ to $T_1$. Furthermore at least $N R^{(T_1)}_i$ 
of those edge-disjoint paths connect source node $i$ (where $i \in
S$) to $T_1$. It follows that if $B\subseteq S$ then the number of
edge disjoint paths from $B$ to $T_1$ is greater than or equal to
$N \sum_{i \in B} R^{(T_1)}_i$.

Now, note that $R^{(T_1)} \in \mathcal{R_{SW}}$ which implies that
for all $B \subseteq S$
\begin{equation*}
N \sum_{i \in B} R^{(T_1)}_i \geq N H(X_B | X_{B^{c}}).
\end{equation*}
This means that there exist at least $ N H(X_B | X_{B^{c}})$
edge-disjoint paths from $B$ to $T_1$ in the subgraph induced by
$T_1$ which in turn implies that $\textrm{min-cut} (B, T_1) \geq N
H(X_B | X_{B^{c}})$ over the subgraph induced by $x^{(T_1)}$. This
holds for all $T_k \in T$, as we have a feasible $x^{(T_k)}$ for
all the terminals. Finally $z$ induces a subgraph where this
property continues to hold true for each terminal since
$x^{(T_k)}_{ij} \leq z_{ij}$, for all $(i,j) \in E^*, T_k \in T$.
Therefore for each terminal $T_k$ we have shown that
$\textrm{min-cut} (B, T_k) \geq N H(X_B | X_{B^{c}})$ over
$G^{*N}_z$ for all sufficiently large $N$. Using Theorem \ref{thm:
tracey_ciss_thm} we have the required proof.
\endproof


The formulation of {\it MIN-COST-SW-NETWORK} as presented above is
a linear program and can potentially be solved by a regular LP
solver. However the number of constraints due to the requirement
that $\mathbf{R} \in \mathcal{R_{SW}}$ is $|T|(2^{N_S} - 1)$ that
grows exponentially with the number of sources. For regular LP
solvers the time complexity scales with the number of constraints
and variables. Thus, using a regular LP solver is certainly not
time-efficient. Moreover even storing the constraints consumes
exponential space and thus using a regular LP solver would also be
space-inefficient. In the sequel we present efficient techniques
for solving this problem.
\subsection{Solving  MIN-COST-SW-NETWORK via dual decomposition}
Suppose that we are given an instance of the S-W problem over a
network specified by $P = <R_{SW}, G, S, T>$. We assume that $P$
is feasible. The {\it MIN-COST-SW-NETWORK} optimization problem is
a linear program and therefore feasibility implies that strong
duality holds \cite{boyd_van}.

We shall refer to the variables $z$, $x^{(T_k)}, R^{(T_k)}$ for
$T_k \in T$ as the primal variables. To simplify notation we let
$x_{s^*}^{(T_k)} = [ x_{s^* 1}^{(T_k)}~ x_{s^* 2}^{(T_k)}~ \dots~
x_{s^* N_S}^{(T_k)}]^T$ denote the vector of flow variables
corresponding to terminal $T_k$ on the edges from the virtual
super node $s^*$ to the source nodes in $S$. We form the
Lagrangian of the optimization problem with respect to the
constraints $R_i^{(T_k)} \leq x_{s^* i}^{(T_k)}, \text{~~for~} i
\in S, T_k \in T$. This is given by
\begin{equation*}
\begin{split}
&L(\lambda, z,x^{(T_1)}, \dots, x^{(T_{N_R})}, R^{(T_1)}, \dots,
R^{(T_{N_R})}) \\
&= f^T z + \sum_{k=1}^{N_R} \mathbf{\lambda}^T_{k} ( R^{(T_k)} -
x_{s^*}^{(T_k)}),
\end{split}
\end{equation*}
where $\lambda = [\lambda_1^T ~ \lambda_2^T ~ \dots ~
\lambda_{N_R}^T]^T$ is the dual variable such that  $\lambda
\succeq 0$ (where $\succeq$ denotes component-wise inequality).

\par For a given $\lambda$, let $g(\lambda)$ denote the dual
function obtained by minimizing $L(\lambda, z,x^{(T_1)}, \dots,
x^{(T_{N_R})}, R^{(T_1)}, \dots, R^{(T_{N_R})})$ over
$z,x^{(T_1)}, \dots, x^{(T_{N_R})}, R^{(T_1)}, \dots,
R^{(T_{N_R})}$.
Since strong duality holds in our problem we are guaranteed that
the optimal value of {\it MIN-COST-SW-NETWORK} can be equivalently
found by maximizing $g(\lambda)$ subject to $\lambda \succeq 0$
\cite{boyd_van}. Thus, if $g(\lambda)$ can be determined in an
efficient manner for a given $\lambda$ then we can hope to solve
{\it MIN-COST-SW-NETWORK} efficiently.
\par Consider the optimization problem for a given $\lambda
\succeq 0$.
\begin{equation*}
\begin{split}
&\text{minimize~~} f^{T} z + \sum_{k = 1}^{N_R}
\mathbf{\lambda}_{k}^{T} ( R^{(T_k)} - x_{s^*}^{(T_k)})  \\
&\text{s. t.~~} 0 \leq x_{ij}^{(T_k)} \leq z_{ij} \leq C_{ij} , \text{~~} (i,j) \in E^{*}, T_k \in T \\
&\sum_{\{j|(i,j) \in E^{*}\}} x_{ij}^{(T_k)} - \sum_{\{j | (j,i)
\in
E^{*}\}} x_{ji}^{(T_k)} = \sigma_i^{(T_k)}, ~ i \in V^{*}, T_k \in T \\
& R^{(T_k)} \in \mathcal{R_{SW}},  ~T_k \in T.
%
%
%
%
%
%
%
%
%
%
%
\end{split}
\end{equation*}

\par We realize on inspection that
this minimization decomposes into a set of independent subproblems
shown below.
\begin{equation}
\label{eq:SW_decomp_lp_multicast}
\begin{split}
&\text{minimize~~} f^{T}z - \sum_{k = 1}^{N_R}
\lambda_{k}^{T} x_{s^*}^{(T_k)}  \\
&\text{s. t.~~} 0 \leq x_{ij}^{(T_k)} \leq z_{ij} \leq C_{ij} , \text{~~} (i,j) \in E^{*}, T_k \in T \\
&\sum_{\{j|(i,j) \in E^{*}\}} x_{ij}^{(T_k)} - \sum_{\{j | (j,i)
\in
E^{*}\}} x_{ji}^{(T_k)} = \sigma_i^{(T_k)}, ~ i \in V^{*}, T_k \in T \\
\end{split}
\end{equation}
and for each $T_k \in T$,
\begin{equation}
\label{eq:SW_decomp_contra}
\begin{split}
&\text{minimize~~} \lambda_k^{T} R^{(T_k)}\\
&\text{subject to~~} R^{(T_k)} \in \mathcal{R_{SW}}.
\end{split}
\end{equation}

The optimization problem in (\ref{eq:SW_decomp_lp_multicast}) is a
linear program with variables $z$ and $x^{(T_k)}$ for $k = 1,
\dots, N_R$ and a total of $(2 |T| + 1)|E^*| + |T||V^*|$
constraints that can be solved efficiently by using a regular LP
solver. It can also be solved by treating it as a minimum cost
network flow problem with fixed rates for which many efficient
techniques have been developed \cite{rk}.

However each of the subproblems in (\ref{eq:SW_decomp_contra})
still has $2^{N_S} - 1$ constraints and therefore the complexity
of using an LP solver is still exponential in $N_S$. However using
the supermodularity property of the conditional entropy function
$H(X_B | X_B^c)$, it can be shown that $\mathcal{R_{SW}}$ is a
contra-polymatroid with rank function $H(X_B | X_{B^c})$
\cite{chen_berger}. In addition, the form of the objective function
is also linear.
It follows that the solution to this problem can be found by a
greedy allocation of the rates as shown in \cite{tsehanly_1}. We
proceed as follows.
\begin{enumerate}
\item Find a permutation $\pi$ such that $\lambda_{k, \pi(1)} \geq
\lambda_{k, \pi(2)} \geq \dots\geq \lambda_{k, \pi(N_S)}$. \item
Set
\begin{eqnarray}
\label{greedy_soln}
\begin{split}
R^{(T_k)}_{\pi(1)} &= H(X_{\{\pi(1)\}} | X_{\{\pi(1)\}^c})
\text{~~and}\\
R^{(T_k)}_{\pi(i)} &= H(X_{\{\pi(1), \dots, \pi(i)\}} |
X_{\{\pi(1), \dots, \pi(i)\}^c})\\
 &- H(X_{\{\pi(1), \dots,
\pi(i-1)\}}| X_{\{\pi(1), \dots, \pi(i-1)\}^c})\\
&\text{~for $2 \leq i \leq N_S$.}
\end{split}
\end{eqnarray}
\end{enumerate}

The previous algorithm presents us a technique for finding the
value of $g(\lambda)$ efficiently. It remains to solve the
maximization
\begin{equation*}
\max_{\lambda \succeq 0} g(\lambda).
\end{equation*}
For this purpose we use the fact that the dual function is concave
(possibly non-differentiable) and can therefore be maximized by
using the projected subgradient algorithm \cite{nonlinear_bert}.
The subgradient for $\lambda_{k}$ can be found as $R^{(T_k)} -
x_{s^*}^{(T_k)}$ \cite{nonlinear_bert}.

\par Let $\lambda^{i}$ represent the value of the dual variable
$\lambda$ at the $i^{th}$ iteration and $\theta_i$ be the step
size at the $i^{th}$ iteration. A step by step algorithm to solve
{\it MIN-COST-SW-NETWORK} is presented below.
\begin{enumerate}
\item Initialize $\lambda^0 \succeq 0$. \item For given
$\lambda^i$ solve
\begin{equation*}
\begin{split}
&\text{minimize~~} f^{T}z - \sum_{k = 1}^{N_R}
(\lambda^{i}_{k})^T x_{s^*}^{(T_k)}\\
&\text{s. t.~~} 0 \leq x_{ij}^{(T_k)} \leq z_{ij} \leq C_{ij} , \text{~~} (i,j) \in E^{*}, T_k \in T \\
&\sum_{\{j|(i,j) \in E^{*}\}} x_{ij}^{(T_k)} - \sum_{\{j | (j,i)
\in
E^{*}\}} x_{ji}^{(T_k)} = \sigma_i^{(T_k)},\\
& \text{for} ~ i \in V^{*}, T_k \in T \\
\end{split}
\end{equation*}
using an LP solver and for each $T_k \in T$,
\begin{equation}
\label{decomp-SW-subgrad}
\begin{split}
&\text{minimize~~} (\lambda^{i}_{k})^T R^{(T_k)}\\
&\text{subject to~~} R^{(T_k)} \in \mathcal{R_{SW}}
\end{split}
\end{equation}
using the greedy algorithm presented in (\ref{greedy_soln}). \item
Set $\lambda^{i+1}_k = [\lambda^i_k + \theta_i (R^{(T_k)} -
x_{s^*}^{(T_k)})]^{+}$ for all $T_k \in T$. Goto step 2 and repeat
until convergence.
\end{enumerate}

\par While subgradient optimization provides a good approximation
on the optimal value of the primal problem, a primal optimal
solution or even a feasible, near-optimal solution is usually not
available. In our problem, we seek to jointly find the flows and
the rate allocations that support the recovery of the sources at
the terminals at minimum cost. Thus, finding the appropriate flows
and rates specified by the primal-optimal or near primal-optimal
$z, x^{(T_1)},\dots, x^{(T_{N_R})}, R^{(T_1)}, \dots ,
R^{(T_{N_R})}$ is important. Towards this end we use the method of
Sherali and Choi \cite{sheralichoi}.
\par 

We now briefly outline the primal recovery procedure of
\cite{sheralichoi}. Let $\mu_j^k$ for $j = 1, \dots, k$ be a set
of convex combination weights for each $k \geq 1$. This means that
\beq \nonumber \sum_{j = 1}^k \mu_j^k = 1, \text{~and~} \mu_j^k \geq 0. \eeq

We define \beq \nonumber \gamma_{jk} = \mu_j^k / \theta_k, \text{~for~} 1
\leq j \leq k, \text{~and~} k \geq 1,\eeq and let \beq \nonumber
\Delta\gamma_k^{\max} \triangleq \max \{\gamma_{jk} -
\gamma_{(j-1)k}: j = 2, \dots, k \}.\eeq

Let the primal solution returned by subgradient optimization at
iteration $k$ be denoted by the vector $(z, x, R)_{k}$ and let the
$k^{th}$ primal iterate be defined as \beq (\tilde{z}, \tilde{x},
\tilde{R})_{k} = \sum_{j=1}^k \mu_{j}^k  (z, x, R)_{j} \text{~~for
$k \geq 1$}.\eeq

Suppose that the sequence of weights $\mu_j^k$ for $k \geq 1$ and
the sequence of step sizes $\theta_k, k \geq 1$ are chosen such
that

\begin{enumerate}
\item $\gamma_{jk} \geq \gamma_{(j-1)k}$ for all $j = 2, \dots, k$
for each $k$. \item $\Delta\gamma_k^{\max} \rightarrow 0$, as $k
\rightarrow \infty$, and \item $\gamma_{1k} \rightarrow 0$ as $k
\rightarrow \infty$ and $\gamma_{kk} \leq \delta$ for all $k$, for
some $\delta > 0$.
\end{enumerate}
Then an optimal solution to the primal problem can be obtained
from any accumulation point of the sequence of primal iterates
$\{(\tilde{z}, \tilde{x}, \tilde{R})\}$.
\par Some useful choices for the step sizes $\theta_k$ and the
convex combination weights $\mu_j^k$ that satisfy these conditions
are given below (see \cite{sheralichoi}).
\begin{enumerate}
\item $\theta_k = a/(b + ck)$,  for $k \geq 1$ where $a > 0, b
\geq 0$, and $c \geq 0$ and $\mu_j^k = 1/k$ for all $j = 1, \dots,
k$. \item $\theta_k = k^{-\alpha}$, for $k \geq 1$ where $0 <
\alpha < 1$ and $\mu_j^k = 1/k$ for all $j = 1, \dots, k$.
\end{enumerate}

\par The strategy for obtaining a near-optimal primal solution for the {\it MIN-COST-SW-NETWORK} problem is
now quite clear. We run the subgradient algorithm in the manner
outlined above and keep computing the sequence of primal iterates
$\{(\mathbf{\tilde{z}}, \mathbf{\tilde{x}},
\mathbf{\tilde{R}})\}_{k \geq 1}$ and stop when the primal
iterates have converged. 

\subsection{Results}
In this section we present results on the performance of our
proposed algorithm. We generated graphs at random by choosing the
position of the nodes uniformly at randomly from the unit square.
Two nodes were connected with an edge of capacity $40.0$ if they
were within a distance of $\frac{0.3}{\sqrt{2}}$ of each other and
were connected with an edge of capacity $20.0$ if they were within
a distance of $0.3$ of each other. The orientation of the edges is
from left to right. A certain number of nodes were declared to be
sources, a certain number to be terminals and the remaining nodes
were used for relaying.

Let the random vector at the sources be denoted by $\mathbf{Z} =
(Z_1, Z_2, \dots, Z_{N_S})$. As in \cite{razvan}, a jointly
Gaussian model was assumed for the data sensed at the sources.
Thus the pdf of the observations is assumed to be \beqno
\begin{split}
f(z_1, z_2, \dots, z_{N_S}) &= \frac{1}{\sqrt{2\pi}^{N_S}
\sqrt{\det (C_{ZZ})}} \\
&\times \exp{\big{(}- \frac{1}{2} (\mathbf{z} - \mu)^T C_{ZZ}^{-1}
(\mathbf{z} - \mu)\big{)}}, \end{split}
 \eeqno where $C_{ZZ}$ is the covariance of the observations.
 We assumed a correlation model where $C_{ZZ}(i,i) = \sigma^2_i$
 and $C_{ZZ}(i,j) = \sigma^2 \exp( - c d_{ij}^{\beta})$ when $i \neq j$ (where $c$
 and $\beta$ are positive constants and $d_{ij}$ is the distance between nodes $i$ and
 $j$). It is further assumed that the samples are quantized independently at all source
nodes with the same quantization step $\Delta$ that is
sufficiently small. Under these conditions, the quantized random
vector $\mathbf{X} = (X_1, X_2, \dots, X_{N_S})$ is such that \beq \nonumber
H(\mathbf{X}) \approx h (\mathbf{Z}) - N_S \log \Delta \eeq as
shown in \cite{coverthomas} where $H(X)$ represents the entropy of
$\mathbf{X}$ and $h(\mathbf{Z}) = \frac{1}{2} \log (2\pi e)^{N_S}
\det (C_{ZZ})$ represents the differential entropy of
$\mathbf{Z}$. We can also express the conditional entropy \beq \nonumber
\begin{split} H(X_B | X_{B^c}) &\approx \frac{1}{2} \log \bigg{(}
(2\pi e)^{N_S - |B^c|} \frac{\det
(C_{ZZ})}{\det (C_{Z_{B^c} Z_{B^c}})} \bigg{)}\\
& - (N_S - |B^c|) \log \Delta \end{split}\eeq We used these
conditional entropies for the Slepian-Wolf region of the sources.

\begin{figure}[htbp]
\subfigure[]{
\includegraphics[width=80mm, clip=true]{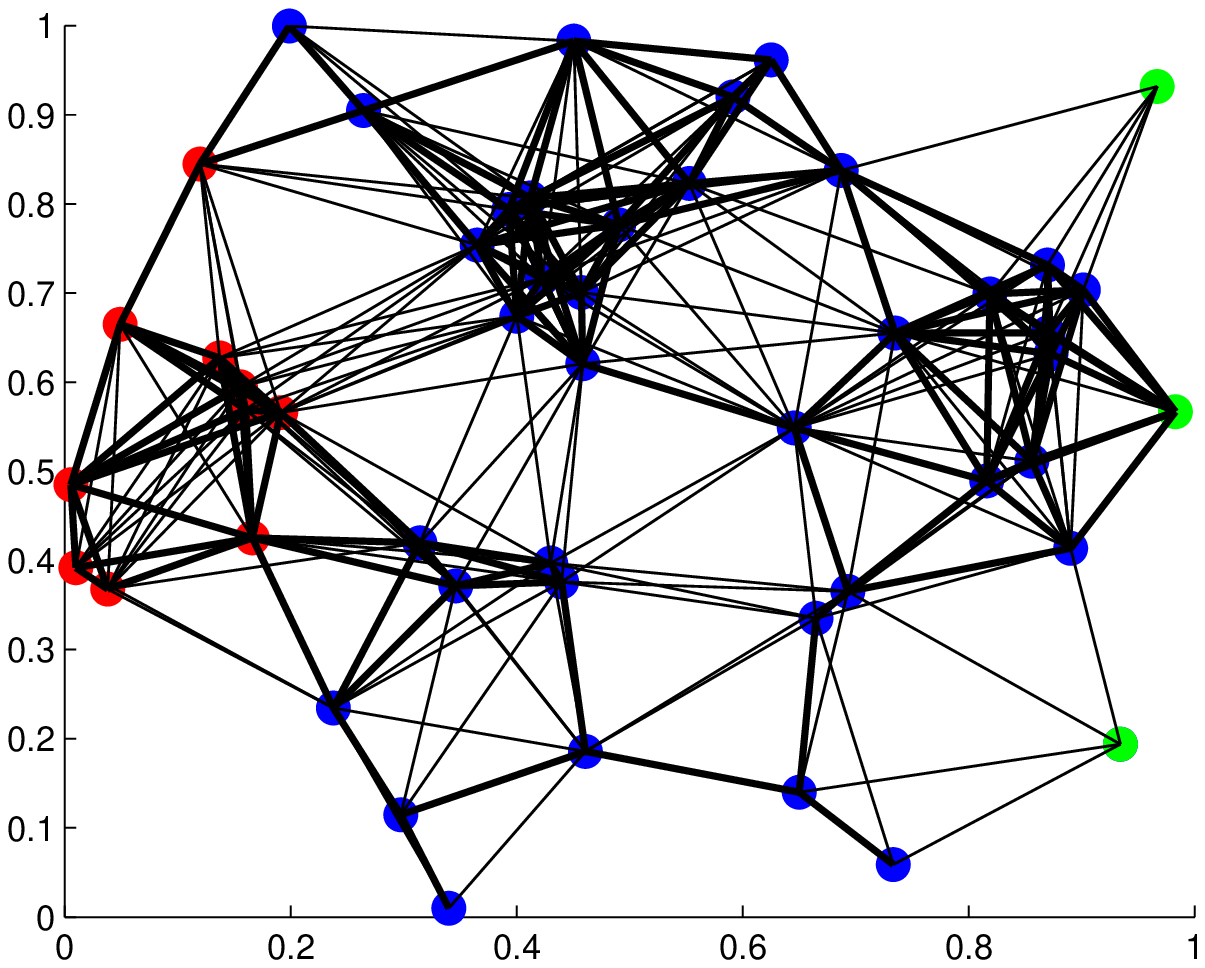}
\label{fig:sw_net_graph}} ~~
\subfigure[]{ \label{fig:sw_subgrad_perf_plot}
\includegraphics[width=80mm, clip=true]{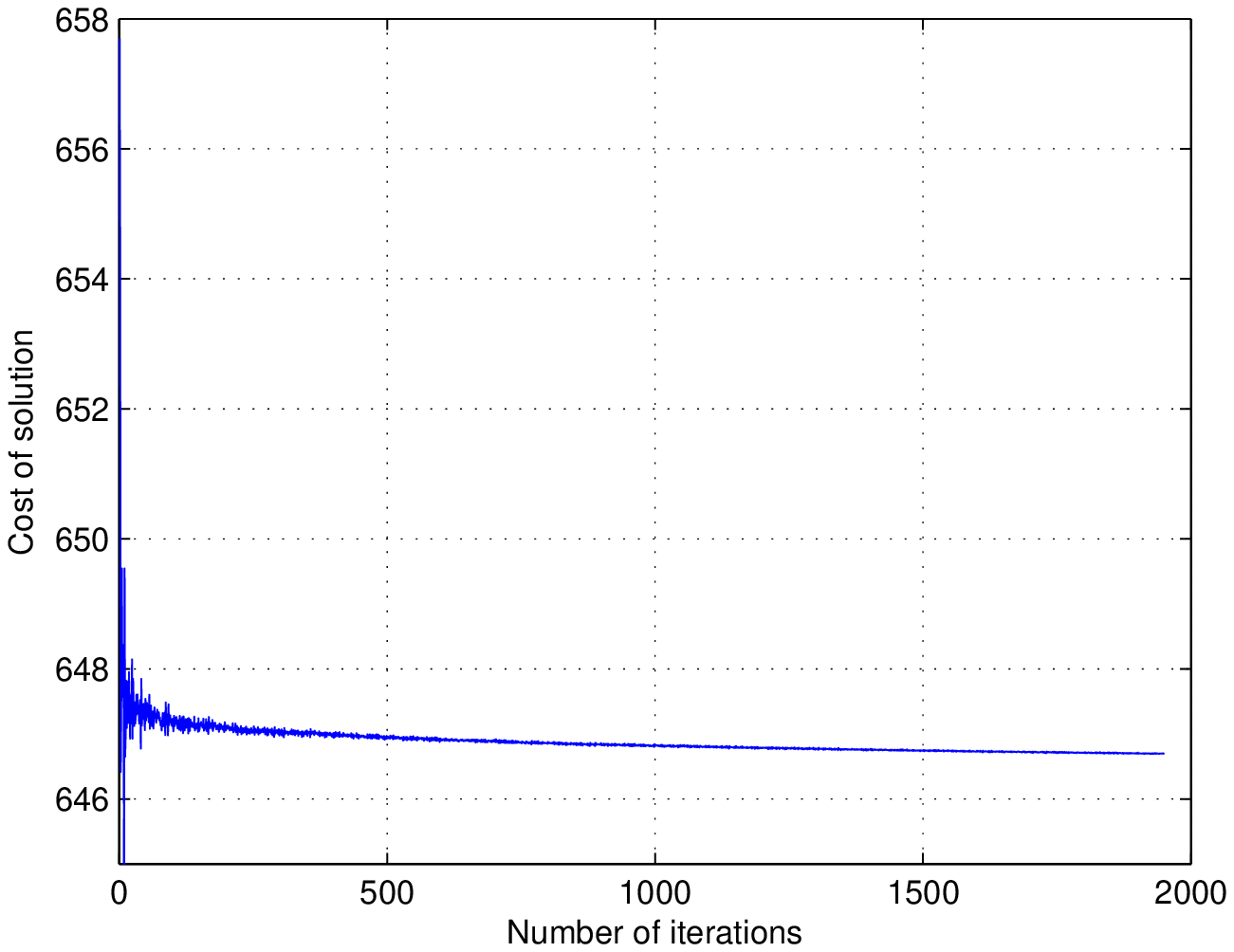}
} \caption{(a) Network with ten source nodes (in red) and three
terminal nodes (in green). The light edges have capacity $20$ and
the dark edges have capacity $40$. All edges are directed from
left to right and have unit cost. (b) Convergence of the
subgradient algorithm to the optimal cost.}
\end{figure}

Figure \ref{fig:sw_net_graph} shows a network consisting of $50$
nodes, with $10$ source nodes, $37$ relay nodes and $3$ terminals.
We chose $\sigma^2 = 1$, $c=1$ and $\beta =1$ for this example.
The quantization step size was chosen to be $\Delta = 0.01$. The
cost of all the edges in the graph was set to $1.0$.

For the subgradient algorithm, we chose $\mu_j^k = 1/k $ for all
$j, k$ and the step size $\theta_k = 8/k^{0.8}$. The averaging
process ignored the first 50 primal solution due to their poor
quality. We observe a gradual convergence of the cost of our
solution to the optimal in Fig. \ref{fig:sw_subgrad_perf_plot}.

\subsubsection{Remark 1}
If one uses a regular LP solver to solve the $MIN-COST-SW-NETWORK$ problem, as noted above, the complexity would scale with the number of variables and constraints, that grow exponentially with the number of sources. However, one is guaranteed that the LP solver will terminate in a finite number of steps eventually. Our proposed algorithm uses the subgradient method with step sizes such that the recovered solution will converge to the optimal as the number of iterations go to infinity \cite{nonlinear_bert}. In general, it does not seem possible to claim convergence in a finite number of steps for this method. A discussion around convergence issues of the subgradient method can be found in Chap. 6 of \cite{nonlinear_bert}. We point out that in practice, we found the algorithms to converge well. Note also that even the description of the LP requires space that increases very quickly, therefore using the LP formulation becomes impractical even with a moderate number of sources.

\subsection{Discussion about Fujishige's algorithm}
We now discuss the work of Fujishige \cite{fujishige78}. Towards this end we need to define the following quantities. A polymatroid $P$ is defined as a pair $(A, \rho)$ where $A$ is a finite set and $\rho$ is  a function from $2^A$ to the positive reals, $\calR_{+}$, that satisfies the axioms of a rank function. A vector $\alpha  \in \calR_{+}^{|A|}$, with entries indexed by the elements of $A$ is called an independent vector of $P(A, \rho)$ if $\sum_{e \in B}\alpha(e) \leq \rho(B)$, for all $B \subseteq A$.

Suppose that we have a directed graph $G = (V, E, C)$, with linear costs $f_{ij}, \forall (i,j) \in E$, (as defined above) with two vertex subsets $V_1$ (source nodes) and $V_2$ (terminal nodes). Suppose that for each $V_i, i = 1,2$, a polymatroid $P_i(V_i, \rho_i)$ is defined. An independent flow is a triple $(\alpha, \delta, \beta)$, such that: (i) $\alpha$ is an independent vector of $P_1$ and is the vector of flows entering the network at the source nodes, (ii) $\beta$ is an independent vector of $P_2$ and is the vector of flows absorbed at the terminal nodes, and (iii) $\delta$ is a flow vector such that flow balance is satisfied at each node in $G$. The algorithm in \cite{fujishige78}, returns an independent flow of maximum value, whose cost $f^T\delta$ is minimum.

In the case of a single terminal, this algorithm can be used to solve the $MIN-COST-SW-NETWORK$ problem (as noted in \cite{han80}) as follows. The set of source nodes and the conditional entropy function, specify a contra-polymatroid. An equivalent polymatroidal representation can be found without much difficulty (see \cite{han80}). Thus, these specify $V_1$ and $P_1$. If there is only one terminal, then one can simply define a (trivial) polymatroid on it. This specifies $V_2$ and $P_2$.

The situation is different when one considers network coding and multiple terminals. The algorithm in \cite{fujishige78}, is only guaranteed to return one set of flows that satisfies flow balance at all source nodes, internal nodes and the terminals. It is possible to show the existence of instances where one set of flows will not simultaneously satisfy all the terminals, when one considers the $MIN-COST-SW-NETWORK$ problem. An example can be found in Figure 6 in \cite{adisepDSC}. Moreover, the objective function in $MIN-COST-SW-NETWORK$, penalizes the maximum of the flows across the terminals at each node, which is different from the one in \cite{fujishige78}. Thus, it is unclear whether this algorithm can be adapted to our problem in a straightforward manner.
\section{Quadratic Gaussian CEO over a network}
\label{sec:ceo_over_network} In general, the problem of
transmitting compressible sources over a network need not have the
requirement of lossless reconstruction of the sources. This maybe
due to multiple reasons. The terminal may be satisfied with a low
resolution reconstruction of the sources to save on network
resources or lossless reconstruction may be impossible because of
the nature of sources. If a source is continuous then perfect
reconstruction would theoretically require an infinite number of
bits. Thus the problem of lossy reconstruction has also been an
active area of research. In this section we shall consider the
quadratic Gaussian CEO problem \cite{viswanathanB97} over
a network. We start by outlining the original problem considered
by \cite{viswanathanB97}. We then present the minimum cost
formulation in the network context and present efficient solutions
to it.

Consider a data sequence $\{X(t)\}^{\infty}_{t=1}$ that cannot be
observed directly. Instead, independent corrupted versions of the
data sequence are available at a set of $L$ agents who are in
communication with the Chief Estimation Officer (CEO) over
different communication channels. The agents are not allowed to
cooperate in any fashion. Suppose that the CEO requires the
reconstruction of $\{X(t)\}^{\infty}_{t=1}$ at an average
distortion level of at most $D$. Here, the distortion level is a
metric of the fidelity of the reconstruction. Suppose agent $i$
communicates with the CEO at rate $R_i$. The CEO problem
\cite{bergerZV96} is one of studying the region of feasible rate
vectors $(R_1, R_2, \dots R_L)$ that allow the reconstruction of
the data sequence under the prescribed distortion constraint. As
in the Slepian-Wolf case there is a direct link between the agents
and the terminal (or the CEO). 
The quadratic Gaussian CEO problem is the particular instance of
the CEO problem when the data source $\{X(t)\}^{\infty}_{t=1}$ is
Gaussian and the distortion metric is mean squared error. A formal
description of the problem follows.

Let $\{X(t)\}^{\infty}_{t=1}$ represent a sequence of i.i.d.
Gaussian random variables and $\{Y_i(t)\}^{\infty}_{t=1} = \{X(t)
+ N_i(t)\}^{\infty}_{t=1}, i = 1, \dots, N_S$ where
$\{N_i(t)\}_{t=1}^{\infty}$ are i.i.d. Gaussian independent of
$\{X(t)\}^{\infty}_{t=1}$ with $E(N_i(t)) = 0, Var(N_i(t)) =
\sigma^2_i$. Furthermore $\{N_i(t)\}^{\infty}_{t=1}$ and
$\{N_j(t)\}^{\infty}_{t=1}$ are independent when $i \neq j$.
\par Let $\epsilon > 0$ be a small real number. The $i^{th}$ agent encodes a block of length $n$ from his
observations $\{y_i(t)\}^{\infty}_{t=1}$ (here, $y_i(t)$ denotes a
particular realization of the random variable $Y_i(t)$) using an
encoding function $f^i_n: \mathbb{R}^n \rightarrow \{1, 2, \dots,
\lfloor 2^{ n(R_i + \epsilon)}\rfloor \}$ of rate $R_i +
\epsilon$. The codewords from the $N_S$ sources are sent to the
CEO who seeks to recover an estimate of the source message over
$n$ time instants $(x(1), x(2), \dots, x(n))$ using a decoding
function $g_n:\{1, 2, \dots, \lfloor 2^{ n(R_1 +
\epsilon)}\rfloor\} \times \dots \times \{1, 2, \dots, \lfloor 2^{
n(R_{N_S} + \epsilon)}\rfloor\} \rightarrow \mathbb{R}^n$.

\begin{definition}
A rate vector $(R_1, \dots, R_{N_S})$ is said to be achievable for
a distortion level $D$ if for $\epsilon > 0$, there exists $n_0$
such that for all $n
> n_0$, there exist encoding functions $f^i_n: \mathbb{R}^n
\rightarrow \{1, 2, \dots, \lfloor 2^{ n(R_i + \epsilon)}\rfloor
\}$ and a decoding function $g_n:\{1, 2, \dots, \lfloor 2^{n(R_1 +
\epsilon)} \rfloor\} \times \dots \times \{1, 2, \dots,
 \lfloor 2^{n(R_{N_S} + \epsilon)} \rfloor\} \rightarrow \mathbb{R}^n$ such that
$\frac{1}{n} E \sum_{t=1}^{n} (X(t) - \hat{X}(t))^2 \leq D +
\epsilon$ where $\hat{X}^{n} = g_n(f^1_n(Y^n_1), \dots,
f^{N_S}_n(Y^n_{N_S}))$.
\end{definition}




\par A complete characterization of the feasible rate region for a
given distortion level $D$, denoted by $\mathcal{R(D)}$ has been
obtained in \cite{oohama_1}\cite{prabhatse} and is given
below.


\begin{equation}
\label{eq:quad_gauss_ceo_region} \mathcal{R(D)} = \bigcup_{(r_1,
\dots, r_{N_S}) \in \mathcal{F(D)}} \mathcal{R_D}(r_1, \dots,
r_{N_S})
\end{equation}
where
\begin{equation}
\nonumber
\begin{split}
&\mathcal{R_D}(r_1, \dots, r_{N_S}) \triangleq \bigg{\{}(R_1,
\dots, R_{N_S}) : A \subseteq \{1, \dots, N_S\}, \\
&A\neq \phi,
\sum_{k \in A} R_k \geq \sum_{k \in A} r_k  + \frac{1}{2} \log
\bigg{(}\frac{\frac{1}{\sigma^2_X} + \sum_{r_i = 1}^{N_S} \frac{1-
e^{-2r_i}}{\sigma^2_{i}}}{\frac{1}{\sigma^2_X} + \sum_{i \in A^c}
\frac{1 - e^{-2r_i}}{\sigma^2_i}} \bigg{)}
\bigg{\}}, \\
\end{split}
\end{equation}
\beq \nonumber  \mathcal{F(D)} = \bigg{\{}(r_1, \dots, r_{N_S}):
r_i \geq 0, \frac{1}{\sigma^2_X} + \sum_{i=1}^{N_S} \frac{1 -
e^{-2r_i}}{\sigma^2_i} \geq \frac{1}{D}\bigg{\}}. \eeq

It is important to note that $\mathcal{R(D)}$ is convex
\cite{prabhatse}. Thus, in principle the minimization of a convex
function of $(R_1, R_2, \dots, R_{N_S})$ can be performed
efficiently.

We are interested in the quadratic Gaussian CEO problem over a
network. In line with our general setup presented in 
Section \ref{prob-form-sec}, the $i^{th}$ source node in $S$
observes the process $\{Y_i(t)\}^{\infty}_{t=1} = \{X(t) +
N_i(t)\}^{\infty}_{t=1}$ and encodes the observations at a rate
$R_i$. 
Once again we are interested in the minimum cost network flow
problem with rates such that they permit the recovery of the
source $X$ at the terminals with the desired level of fidelity,
which in this case shall be measured by mean squared error. 

We start by highlighting the differences between this problem and
the minimum cost Slepian-Wolf over a network. In the previous
subsection we observed that the work of Ho et al. shows that
random linear network coding over a subgraph such that $C_{T_i}
\cap \mathcal{R_{SW}} \neq \phi, \forall T_i \in T$ allows the
lossless recovery of the sources at the terminals in $T$ and
essentially the Slepian-Wolf theorem holds even in the network
case with multiple terminals i.e. any rate vector that can be
obtained by joint coding can be obtained by distributed coding
even when there are multiple terminals. However, an analogous
result in the case of the quadratic Gaussian CEO problem does not
exist. Furthermore, the rate region for the quadratic Gaussian CEO
problem over a general network is unknown. As a simple example, we
may have a network where two source nodes are connected to a
common intermediate node. The intermediate node can then combine
the quantized observations from these source nodes to generate a
new quantized observation such that a lower rate is possible. Thus
the rate region given by the classical Gaussian CEO problem may
not hold as the two codewords may be fused to produce a new
codeword that enables lower rate transmission. 

The first issue can be handled by assuming that there is only one
terminal, i.e., $N_R = 1$ and $C_{T_1} \cap \mathcal{R(D)} \neq
\phi$ so that routing will suffice to transmit a rate vector
belonging to $C_{T_1} \cap \mathcal{R(D)}$ to the terminal $T_1$.
Thus in this problem we shall not consider network coding. For the
second issue we assume that the network operates in a separate
compression and information transfer mode. The set of source nodes
quantize their observations as they would in the original
quadratic Gaussian CEO problem. After this source coding step, the
network ignores any data correlations and routes the data as
though it were incompressible. In general this separation of the
compression and the information transfer is suboptimal, however it
is likely to be a simple way of operating the network.

It is more convenient to cast this optimization in terms of the
original graph rather than the augmented graph. The {\it
MIN-COST-QUAD-CEO-NETWORK} problem becomes
\begin{equation}
\nonumber
\begin{split}
& \text{minimize~} \sum_{(i,j) \in E} f_{ij}x_{ij} \\
&\text{subject to~~} 0 \leq x_{ij} \leq C_{ij} , \text{~~~} (i,j) \in E\\
\end{split}
\end{equation}
\begin{align}
\label{eq:min-cost-ceo-network-fb1}\sum_{\{j|(i,j) \in E\}} x_{ij}
- \sum_{\{j | (j,i) \in
E\}} x_{ji} &= 0,i \in (S \cup \{T_1\})^c\\ 
\label{eq:min-cost-ceo-network-fb2} \sum_{\{j|(i,j) \in E\}}
x_{ij} - \sum_{\{j | (j,i) \in
E\}} x_{ji} &\geq R_i , ~ i \in S \\
\label{eq:min-cost-ceo-network-fb3} \sum_{\{j|(i,j) \in E\}}
x_{ij} - \sum_{\{j | (j,i) \in
E\}} x_{ji} &\leq - \sum_{i \in S} R_i,~ i = T_1\\
R &\in \mathcal{R(D)} \nonumber
\end{align}
Here (\ref{eq:min-cost-ceo-network-fb1}) enforces the flow balance
at all nodes in $V$ except those in $S \cup \{T_1\}$,
(\ref{eq:min-cost-ceo-network-fb2}) enforces the constraint that
at least $R_i$ units of flow is injected at each source node $i
\in S$ and (\ref{eq:min-cost-ceo-network-fb3}) ensures that at
least $\sum_{i \in S} R_i$ is received at the terminal $T_1$. For
the {\it MIN-COST-SW-NETWORK} problem the total rate to be
transmitted to $T_1$ could be fixed to $H(X_1, \dots, X_N)$ as
shown in the Appendix. However for the problem presented above
fixing the total rate is not possible because of the nature of the
inequalities specifying $\mathcal{R(D)}$. A feasible solution to
the optimization presented above would yield a routing solution
such that the delivery of a rate vector belonging to
$\mathcal{R(D)}$ is possible at terminal $T_1$. The proof is
similar to the one presented in the proof of Lemma
\ref{feas-lemma}. However even though the optimization under
consideration above is convex, the number of constraints
specifying $\mathcal{R(D)}$ is exponential in $N_S$ that would
make a regular convex program solver inefficient.

\subsection{Solving {\it MIN-COST-QUAD-CEO-NETWORK} via dual decomposition}
We assume that the {\it MIN-COST-QUAD-CEO-NETWORK} problem is
strictly feasible so that
strong duality holds \cite{boyd_van}. 
The Lagrangian with respect to the set of flow balance constraints
that contain terms dependent on $R_i, i \in S$ for a given
$\lambda$ is given by
\begin{equation}
\nonumber
\begin{split}
&L(\lambda,\mathbf{x},R) = \mathbf{f}^T \mathbf{x} + \sum_{i \in
S} \lambda_i \bigg{(}R_i - \sum_{\{j|(i,j) \in E\}} x_{ij}\\
&+ \sum_{\{j | (j,i) \in E\}} x_{ji} \bigg{)} + \lambda_{T_1}
\bigg{(} \sum_{i \in S} R_i + \sum_{\{j|(T_1,j) \in E\}}
x_{T_1j}\\
&- \sum_{\{j | (j,T_1) \in E\}} x_{jT_1} \bigg{)}.
\end{split}
\end{equation}
It is easy to see that finding the dual function $g(\lambda) =
\min_{x, R} L(\lambda, x, R)$ subject to the remaining constraints
decomposes as
\begin{equation}
\label{decomp-lp-ceo}
\begin{split}
&\text{minimize~~} f^{T} x - \sum_{i \in S} \lambda_i
\bigg{(}\sum_{\{j|(i,j) \in E\}} x_{ij} - \sum_{\{j | (j,i) \in
E\}} x_{ji}\bigg{)} \\
&+ \lambda_{T_1} \bigg{(}\sum_{\{j|(T_1,j) \in
E\}} x_{T_1 j} - \sum_{\{j |(j,T_1) \in E\}} x_{j T_1}\bigg{)} \\
&\text{subject to~~} 0 \leq x_{ij} \leq C_{ij} ,
\text{~~~}\forall~
(i,j) \in E,\\
&\sum_{\{j|(i,j) \in E\}} x_{ij} - \sum_{\{j | (j,i) \in
E\}} x_{ji} = 0 ,  ~~~i \in V \backslash S \cup \{T_1\}, \\
\end{split}
\end{equation}
and
\begin{equation}
\label{decomp-CEO}
\begin{split}
&\text{minimize~~} \sum_{i \in S} \lambda_i R_i + \lambda_{T_1}
\sum_{i\in S} R_i\\
&\text{subject to~~} R \in \mathcal{R(D)}.
\end{split}
\end{equation}
The optimization in (\ref{decomp-lp-ceo}) is a linear program that
can be solved efficiently. To solve (\ref{decomp-CEO}) we note
that for a given $(r_1, \dots, r_{N_S})$ it can be shown that the
region $\mathcal{R_{D}}(r_1, \dots, r_{N_S})$ (defined in
(\ref{eq:quad_gauss_ceo_region})) is a contra-polymatroid
\cite{chen_berger}. Therefore an optimization problem such as
\begin{equation}
\label{opt-contra-ceo}
\begin{split}
&\text{minimize~~} w^{T} R\\
&\text{subject to~~} R \in \mathcal{R_{D}}(r_1, \dots, r_{N_S})
\end{split}
\end{equation}
can be solved in closed form by using the greedy allocation
algorithm presented earlier. Using this fact we show (see
Appendix-II) that the optimization in (\ref{decomp-CEO}) reduces to
a convex optimization problem with $N_S + 1$ constraints that can
be solved efficiently via Lagrange multiplier methods. It is
important to note that the optimal value of the above optimization
problem is $-\infty$ if $w_i < 0$ for any $i$. This is because the
inequalities defining $\mathcal{R(D)}$ do not impose any upper
bounds on the individual rates $R_i$ for $i \in S$. Consequently
the optimization in (\ref{decomp-CEO}) has a finite optimal value
only if $\lambda_i + \lambda_{T_1} \geq 0, \forall i \in S$.

It is clear based on the previous arguments that we can evaluate
$g(\lambda)$ efficiently. We now need to solve the optimization
\beq \nonumber
\begin{split}
\max&~ g(\lambda)\\
\text{subject to~} 
\lambda_i &\geq 0, i \in S, \lambda_{T_1} \geq 0.
\end{split}
\eeq 

For solving this maximization we use the projected subgradient
method \cite{nonlinear_bert}. 
 As noted in Section \ref{sec:sw-over-network} the subgradient
algorithm may not return a primal optimal or primal near-optimal
solution. For primal recovery for the {\it
MIN-COST-QUAD-CEO-NETWORK} problem we use the technique proposed
by Larsson et al. \cite{larsson} that generalizes the method of
\cite{sheralichoi} to the case of general convex programs. We
point out some differences between the two methods below.

The method of Larsson et al. considers an optimization of the
following form. \beqno
\begin{split}
&\min f(x)\\
\text{subject to~}& h_{i} (x) \leq 0, i \in \mathcal{I}\\
&x \in X
\end{split}
\eeqno where the functions $f$ and $h_i, i \in \mathcal{I}$ are
convex and the set $X$ is convex and compact. It assumes the
Slater constraint qualification condition, i.e., the existence of a
strictly feasible point $x_1$ such that $\{x_1 \in X ~|~ h_i(x_1)
< 0, i \in \mathcal{I}\}$ and considers the dual function with
respect to the constraints $h_i, i \in \mathcal{I}$, \beqno
\theta(u) = \min_{x \in X} f(x) + \sum_{i \in \mathcal{I}} u_i
h_i(x) \eeqno and then solves the maximization of the dual
function \beq \nonumber
\begin{split}
\max~ &\theta(u)\\
\text{subject to~}& u \succeq 0
\end{split}
\eeq by using the projected subgradient algorithm. Let $x^{(k)}$
denote the primal solution obtained at the $k^{th}$ iteration.
Section 3.2 of \cite{larsson} shows that for step sizes $\alpha_k
\in \big{[} \frac{\mu}{b+k}, \frac{M}{b+k} \big{]}, k > 0, 0 < \mu
\leq M < \infty, k \geq 0$ the sequence of averages defined by
\beq \label{eq:ceo-avg-rule} \hat{x}^n = \frac{1}{n} \sum_{k =
0}^n x^{(k)} \eeq converges to the primal optimal solution as $n
\goes \infty$. The choice of step sizes is more limited in this
method as compared to \cite{sheralichoi}. In our problem the $h_i$
functions are the linear inequality constraints in
(\ref{eq:min-cost-ceo-network-fb2}) and
(\ref{eq:min-cost-ceo-network-fb3}) and the existence of a
strictly feasible solution is assumed. However, the set
$\mathcal{R(D)}$ is convex but not compact. The condition of
compactness is however rather technical and can be enforced by
imposing a loose upper bound on the rates. In practice, while
running the subgradient algorithm the dual variables $\lambda_i, i
\in S$ and $\lambda_{T_1}$ were constrained to be larger than or
equal to $10^{-10}$ at any iteration to ensure that the optimized
rates were bounded. Averaging the solutions as in
(\ref{eq:ceo-avg-rule}) we observed a steady convergence of the
algorithm to the optimal cost in our simulations.
\subsection{Results}
\begin{figure}[ht]
\subfigure[]{
\includegraphics[width=90mm, clip=true]{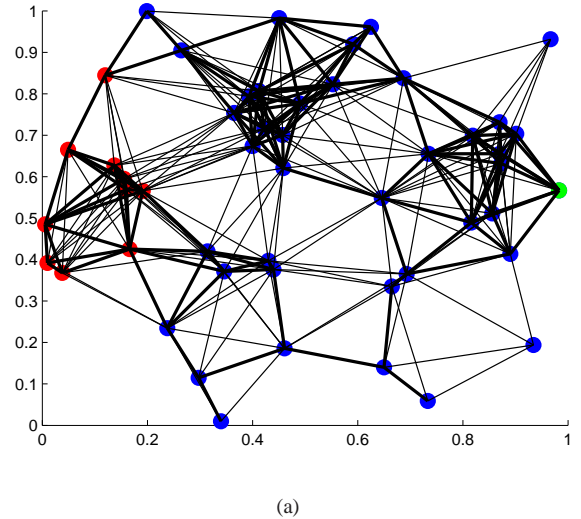}
\label{fig:ceo_net_graph}} ~~
\subfigure[]{ \label{fig:ceo_subgrad_perf_plot}
\includegraphics[width=90mm, clip=true]{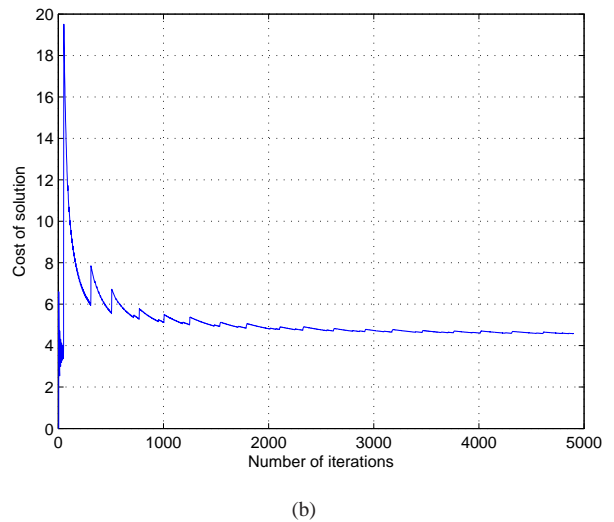}
} \caption{(a) Network with ten source nodes (in red) and one
terminal node (in green). The light edges have capacity $11$ and
the dark edges have capacity $22$. All edges are directed from
left to right and have unit cost. (b) Convergence of the
subgradient algorithm to the optimal cost. }
\end{figure}

As in the previous section we generated the graphs randomly.
However, there is only one terminal in the {\it MIN-COST-QUAD-CEO-NETWORK}
problem since we want a solution based on routing. Figure \ref{fig:ceo_net_graph} shows an example of a
 network with $10$ source nodes, $39$ relay nodes and $1$ terminal
 node. In this particular example we chose the variance of the source to be $\sigma^2_X = 0.01$,
 the sensing noise variance of the source nodes to be $\sigma^2_i =
 0.005$ for all $i \in S$ and the required distortion level $D =
 0.003$. The capacity of the light edges is $11$ and the capacity of the dark edges is $22$.
The cost of each edge was set to 1.0.
  For the subgradient algorithm the step size was chosen to
 be $\alpha_k = \frac{10}{1 + k}$. The averaging
 process ignored the first 100 primal solutions. As demonstrated in Fig.
 \ref{fig:ceo_subgrad_perf_plot} there is a steady convergence of
 the subgradient algorithm to the optimal solution.

\section{Lifetime maximization for sensor networks with data
distortion constraints} \label{sec:lifetime_over_network}

We now consider the problem of maximizing the lifetime of a sensor
network when the terminal node needs to be able to reconstruct the
data at a particular distortion level (related problems were
studied in \cite{kansalramamoorthy}). This problem is important in
the context of sensor networks where nodes are typically
battery-limited and are sensing correlated phenomena that need to
be reconstructed at the terminal node. It has been studied in
\cite{changT00} when the rates of the sensors are fixed. As
explained in the previous section, the sensor nodes observe
independent corrupted versions of an i.i.d. Gaussian data sequence
$\{X(t)\}^{\infty}_{t=1}$ and communicate at a particular rate to
the terminal node. The operating rate vector, whose components
consists of the operating rates of each sensor should be such that
the terminal should be able to reconstruct
$\{X(t)\}^{\infty}_{t=1}$ subject to a mean squared error
distortion constraint. We are interested in finding routes over
which the data should be routed so that the reconstruction (with
an acceptable level of fidelity) can be ensured at the terminal
for the longest period of time before a node runs out of energy.

\begin{figure}[htbp]
\begin{center}
\includegraphics[width=90mm,clip=true]{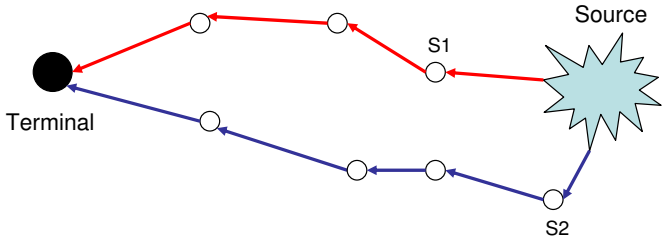}
\caption{\label{fig:lifetime-quad-ceo-eg} The figure shows sensor
nodes $S1$ and $S2$ at different distances from the phenomenon of
interest. The red path corresponds to the data from $S1$ and the
blue path corresponds to the data from $S2$.}
\end{center}
\end{figure}

There is an inherent trade-off between the choice of the operating
rate for a given sensor and the energy consumption that occurs
when the data from the sensor is transmitted to the terminal since
the amount of energy consumption roughly depends on the amount of
distance or hops that the data has to travel.. This is best
illustrated in Fig. \ref{fig:lifetime-quad-ceo-eg}. The sensor
node $S2$ that is closer to the phenomenon of interest has a
better sensing SNR, but is far from the terminal. Therefore $S2$
requires less bits for quantization, however the data needs to
travel a longer distance. On the other hand, sensor node $S1$ that
is further away from the source but closer to the terminal has a
lower sensing SNR and requires more bits for quantization, but its
data needs to travel a smaller distance. 
Thus there is a clear tradeoff in how we would want to perform the
rate and the flow allocation if we wanted to maximize the network
lifetime.

\par In this section we formulate the problem of
maximizing the lifetime of a network in the quadratic Gaussian CEO
setting. As in the previous section we assume that the network
operates by separate compression and network information transfer.
We are given the following.
\begin{itemize}
\item[a)] An i.i.d. Gaussian data source sequence
$\{X(t)\}_{t=1}^{\infty}$ and set of source nodes $S$ that observe
independent corrupted versions of $\{X(t)\}$ given by
$\{Y_i(t)\}_{t=1}^{\infty} = \{X(t) + N_i(t)\}$ for $ i = 1,
\dots, N_S$ where $\{N_i(t)\}_{t=1}^{\infty}$ are i.i.d. Gaussian
independent of $\{X(t)\}^{\infty}_{t=1}$ with $E(N_i(t)) = 0,
Var(N_i(t)) = \sigma^2_i$. There is a single terminal node $T_1$
that seeks to reconstruct the source such that the distortion
under the mean squared error metric is at most $D$. The feasible
rate region for this problem denoted $\mathcal{R(D)}$ is given by
(\ref{eq:quad_gauss_ceo_region}). \item[b)] The initial battery
level, $E_i$ of each node $i \in V$ and the following power
consumption figures of interest.
\begin{itemize}
\item[i)] $P_{tx}(i,j)$ - the power consumed when $i$ transmits
data to $j$ at unit rate for all $i,j \in V$, \item[ii)]
$P_{rx}(i,j)$ - the power consumed when $j$ receives data from $i$
at unit rate for all $i,j \in V$ and, \item[iii)] $P_{sense}(i)$ -
the power consumed when $i$ uses an additional bit for quantizing
its observations.
\end{itemize}
\end{itemize}
We define the network lifetime to be the time until the first node
runs out of energy. The resultant optimization problem that we
call {\it MAX-LIFETIME-DISTORTION-CONSTRAINT} can be expressed as
\begin{equation*}
\begin{split}
&\text{minimize~~} \Gamma\\
&\text{s.t. ~~} 0 \leq x_{ij} \leq C_{ij} , \text{~~~}\forall (i,j) \in E\\
&\sum_{\{j|(i,j) \in E\}} x_{ij} - \sum_{\{j | (j,i) \in
E\}} x_{ji} = 0 , ~ i \in V \backslash S \cup T_1\\
&\sum_{\{j|(i,j) \in E\}} x_{ij} - \sum_{\{j | (j,i) \in
E\}} x_{ji} \geq R_i , ~ i \in  S \\
&\sum_{\{j|(i,j) \in E\}} x_{ij} - \sum_{\{j | (j,i) \in
E\}} x_{ji} \leq - \sum_{j \in S} R_j , ~ i = T_1\\
&\sum_{\{j|(i,j) \in E\}} P_{tx}(i,j) x_{ij} + \sum_{\{j | (j,i)
\in E\}} P_{rx}(j,i) x_{ji} \\
&+ P_{sense} R_i \leq  E_i \Gamma, ~i
\in S\\
&\sum_{\{j|(i,j) \in E\}} P_{tx}(i,j) x_{ij}\\ &+ \sum_{\{j |
(j,i) \in E\}} P_{rx}(j,i) x_{ji} \leq  E_i \Gamma, ~ i \in V
\backslash S\\
&R \in \mathcal{R(D)}
\end{split}
\end{equation*}
where $\Gamma$ denotes the reciprocal of the network lifetime.
Note that \cite{changT00} considers the lifetime maximization
problem when the operating rate for each node $i \in V$ is fixed.
Therefore the problem becomes a linear program that can be solved
efficiently. However we are interested in jointly optimizing the
operating rates and the lifetime of the network. In the
formulation above we note that the specification of the region
$\mathcal{R(D)}$ is non-linear (although convex) with
exponentially (in the number of sources) many inequalities. In
\cite{kansalramamoorthy}, the authors considered a problem similar
to {\it MAX-LIFETIME-DISTORTION-CONSTRAINT} and proposed
suboptimal solutions for  by approximating the constraints of
$\mathcal{R(D)}$ by linear inequalities. The authors presented
approximate linear programs that were obtained by strengthening
and weakening these constraints and concluded that the true
network lifetime was between the results obtained by solving these
linear programs. However the number of constraints was still
exponential in the number of sources that precluded solving large
instances of the problem. In this section we present a solution to
this problem based on dual decomposition.

We point out that we are considering a strategy where the rate
allocation is static. In practice, it may be beneficial to adapt
the rate allocation over time to extend the lifetime of the
network. 
\subsection{Solving {\it MAX-LIFETIME-DISTORTION-CONSTRAINT} by dual decomposition}
 We note that as in section \ref{sec:ceo_over_network} we can dualize
the appropriate flow balance and energy consumption constraints
and compute the dual function for this problem efficiently by
exploiting the contra-polymatroidal structure of $\mathcal{R(D)}$.
However, in practice we observed that the simple projected
subgradient algorithm for maximizing the dual function for this
problem is far too slow to be practical. Therefore we pursue an
alternate line of attack here. We actually minimize $\Gamma^2$
instead of $\Gamma$ and use the proximal bundle method
\cite{kiwiel95} and perform primal recovery as explained in
\cite{kiwiel95}. The Lagrangian for a given $\lambda_1$ and
$\lambda_2$ becomes \beq \nonumber
\begin{split}
&L(\lambda_1,\lambda_2, x, R) = \\
&\Gamma^2 + \sum_{i \in S}
\lambda_{1i} \bigg{(} R_i - \sum_{\{j|(i,j) \in E\}} x_{ij} +
\sum_{\{j | (j,i) \in E\}} x_{ji} \bigg{)}\\ &+ \lambda_{1T_1}
\bigg{(} \sum_{k \in S} R_k + \sum_{\{j|(T_1,j) \in E\}} x_{T_1j}
- \sum_{\{j | (j,T_1)
\in E\}} x_{jT_1} \bigg{)}\\
&+ \sum_{i \in S} \lambda_{2i} \bigg{(} \sum_{\{j|(i,j) \in E\}}
P_{tx}(i,j) x_{ij} + \sum_{\{j | (j,i) \in E\}} P_{rx}(j,i)
x_{ji}\\
&+ P_{sense} R_i - E_i \Gamma \bigg{)}
\end{split}
\eeq and finding $g(\lambda_1, \lambda_2) = \min_{\Gamma, x, R}
L(\lambda_1,\lambda_2, x, R)$ subject to the remaining constraints
decomposes as 
\begin{align}
&\text{minimize~} \Gamma^2  - \Gamma \sum_{i \in S} \lambda_{2i}
E_i \nonumber\\
& + \sum_{i \in S} \lambda_{1i} \bigg{(}- \sum_{ \{j | (i,j)
\in E \}} x_{ij} + \sum_{\{j | (j,i) \in E \}} x_{ji}\bigg{)} \nonumber\\
&+ \lambda_{1 T_1} \bigg{(} \sum_{ \{j | (T_1,j) \in E \}} x_{T_1
j} - \sum_{\{j | (j,T_1) \in E \}} x_{jT_1} \bigg{)} \nonumber\\
&+ \sum_{i\in S} \lambda_{2i} \bigg{(}\sum_{\{j|(i,j) \in E\}}
P_{tx}(i,j) x_{ij} + \sum_{\{j | (j,i) \in E\}} P_{rx}(j,i) x_{ji}\bigg{)} \nonumber\\
&\text{subject to~~} 0 \leq x_{ij} \leq C_{ij} , \text{~~}\forall (i,j) \in E \nonumber\\
&\sum_{\{j|(i,j) \in E\}} x_{ij} - \sum_{\{j | (j,i) \in
E\}} x_{ji} = 0 , ~ i \in V \backslash S \cup T_1 \nonumber\\
&\sum_{\{j|(i,j) \in E\}} P_{tx}(i,j) x_{ij} + \sum_{\{j | (j,i)
\in E\}} P_{rx}(j,i) x_{ji} \leq  E_i \Gamma, ~ i
\in V \backslash S \nonumber\\
\label{eq:max-lifetime-gamma-decomp}
\end{align}
and \beq \label{eq:max-lifetime-rate-decomp}
\begin{split}
\text{minimize~} & \sum_{i \in S} (\lambda_{1i} + \lambda_{1T_1} + P_{sense} \lambda_{2i}) R_i \\
\text{subject to ~} &R \in \mathcal{R(D)}.
\end{split}
\eeq The optimization problem in
(\ref{eq:max-lifetime-gamma-decomp}) can be solved by using a
quadratic programming solver and the optimization problem in
(\ref{eq:max-lifetime-rate-decomp}) can be solved as shown in
Section \ref{sec:ceo_over_network}. We used the quadratic
programming package offered by MOSEK
\footnote{http://www.mosek.com} for this part of the work.

From the above decomposition we have an efficient method to
evaluate $g(\lambda_1, \lambda_2)$. It remains to evaluate \beq \nonumber
\begin{split}
\text{maximize~} &g(\lambda_1, \lambda_2)\\
\text{subject to~} & \lambda_1, \lambda_2 \succeq 0\\
\end{split}
\eeq for which we used the method of \cite{kiwiel95}. We now
briefly overview the proximal bundle technique.

Consider the convex optimization problem \beq \nonumber
\begin{split}
\min ~&\psi_0(z)\\
\text{subject to}~ &\psi_j(z) \leq 0, j = 1 \dots n\\
&z \in Z
\end{split}
\eeq where $Z$ is a compact and convex set, $\psi_j$ is a convex
function for $j = 0 \dots n$. Let $f$ be the dual function of this
optimization problem with respect to the constraints $\psi_j(z)
\leq 0, j = 1 \dots n$. Note that the dual function is always
concave and that we may not know $f$ in its functional form.
However we assume $f$ and a subgradient of $f$ at any given point
is available via an oracle. We are interested in solving the
original convex optimization by finding $\max_{x \in A} f(x)$
where $A$ is a non-empty, closed
convex set and performing primal recovery. Toward this end we use the following algorithm.\\
{\bf Proximal Bundle Algorithm}
\begin{itemize}
\item {\bf Step 1:} Let $\bar{\delta} > 0, m \in (0,1)$. Choose an
initial point $\hat{x}^0$, set $y^0 = \hat{x}^0,$ and let $k = 0$.
Compute $f(\hat{x}^0)$ and a subgradient $s_0$ at $\hat{x}^0$.
Define the $0^{th}$ polyhedral approximation to $f$ as
$\hat{f}_0(y) = f(\hat{x}^0) + s_{0}^{T}(y - \hat{x}^0)$. \item
{\bf Step 2:} At the $k^{th}$ iteration, compute \beq \nonumber y^{k+1} \in
\arg \max_{y \in A} \bigg{[} \hat{f}_k(y) - \frac{\mu_k}{2}
\text{\textdoublevertline} y - \hat{x}^k
\text{\textdoublevertline}^2 \bigg{]},\eeq where $\mu_k$ is a
proximity weight. Store the Lagrange multipliers corresponding to
this optimization denoted by $\nu_j^k \geq 0, 1 \leq j \leq k$
such that $\sum_j \nu_j^k = 1$. \item {\bf Step 3:} 
Define $\delta_k = \hat{f}_k(y) - f(\hat{x}^k)$. If $\delta_k <
\bar{\delta}$ STOP. \item {\bf Step 4:} Compute $f(y^{k+1})$ and a
subgradient $s_{k+1}$ at $y^{k+1}$. Also store the value of the
primal variables corresponding to $y^{k+1}$, denoted by
$z^{k+1}$.\item {\bf Step 5:} If $f(y^{k+1}) - f(\hat{x}^k) \geq m
\delta_k$, perform a $SERIOUS~ STEP~ (SS) ~\hat{x}^{k+1} =
y^{k+1}$, else perform a $NULL~ STEP ~(NS) ~\hat{x}^{k+1} =
\hat{x}^k$. \item {\bf Step 6:} Update the model \beq \nonumber
\hat{f}_{k+1}(y) = \min \{ \hat{f}_k(y), f(y^{k+1}) + s_{k+1}^T(y
- y^{k+1}) \}. \eeq \item {\bf Step 7:} Set $k = k+1$ and goto
Step 2
\end{itemize}

The work of \cite{kiwiel95} shows that the aggregate primal
solution obtained by computing $\sum_{j=1}^k \nu_j^k z^j$ produces
an asymptotically optimal primal solution as $k \goes \infty$. For
more details (such as the choice of the $\mu_k$ sequence) and for techniques for reducing the storage
requirements associated with this method, we refer the reader to
\cite{kiwiel95}.

We applied this method to our problem since we can efficiently
evaluate the dual function and can compute a subgradient at each
point as well. Note that in this method as well, there is the
technical compactness condition on $Z$. In our problem since there
are no upper bounds on the rates, our region is not compact.
However as in the previous section we impose loose upper bounds on
the rates by enforcing the dual variables to be larger than or
equal to $10^{-10}$.
\subsection{Results}
\begin{figure}[htbp]
\begin{center}
\includegraphics[width=80mm,clip=true]{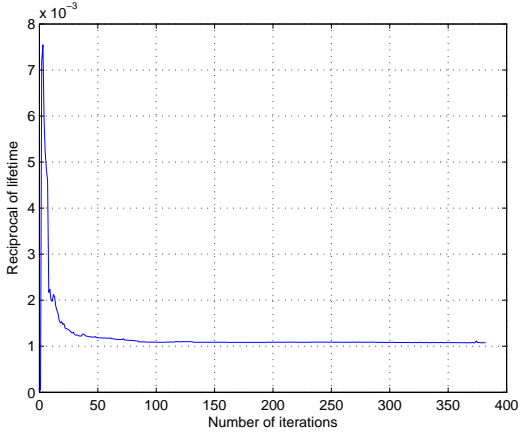}
\caption{\label{fig:lifetime_cutting_plane_perf_plot} Convergence
of the proximal bundle algorithm to the optimal value of
$\Gamma$.}
\end{center}
\end{figure}

We ran the previous algorithm on the same topology shown in Fig.
\ref{fig:ceo_net_graph}. We chose the variance of the source to be
$\sigma^2_X = 0.01$, the sensing noise variance of the source
nodes to be $\sigma^2_i = 0.005$ for all $i \in S$ and the
required distortion level $D = 0.003$. The battery levels for all
the nodes were chosen to be $200$ and we set $P_{tx}(i,j) = 1.0,
P_{rx}(i,j) = 0.5$ and $P_{sense} = 0.001$. As demonstrated in
Fig. \ref{fig:lifetime_cutting_plane_perf_plot} there is a steady
convergence of the proximal bundle algorithm to the optimal
solution.

We note that the lifetime maximization problem can also be solved in a similar manner if one consider the case of lossless reconstruction with multiple terminals.
\section{Conclusions and Future Work}
\label{sec:conclusions} We considered the problem of jointly
allocating rates and flows at minimum cost for distributed source
coding problems over a network. In particular, we considered (a)
the Slepian-Wolf problem, (b) the minimum cost quadratic Gaussian
CEO problem and (c) the problem of maximizing network lifetime
when a source needs to be reconstructed within a quadratic
distortion constraint. These problems are of interest in domains
such as sensor networks where the data that is sensed by different
sensors is typically highly correlated. The feasible rate region of distributed source coding problems is typically
specified by a number of inequalities that is exponential in the
number of sources that makes these problems hard to solve. We
presented an approach based on dual decomposition that uses the
special structure of the rate regions to efficiently compute the
dual function. Finally, we demonstrated approaches for maximizing
the dual function using the subgradient algorithm and the proximal
bundle algorithm.

It would be interesting to investigate algorithms along the lines of those considered by Fujishige \cite{fujishige78} (that do not use dual decomposition), for the problems considered in this paper and study whether they have lower complexity.
In all the problems we considered, we were able to decouple the rate allocations from the flow allocations. This essentially happens because the two sets of variables have a limited interaction via simple linear inequalities. As pointed out by a reviewer there may be other problems that may lend themselves to this kind of decomposition, where the interaction between these variables is more complex. 
\section{Acknowledgements}
The author would like to thank the anonymous reviewers whose suggestions greatly improved the quality and presentation of the paper.
\section{Appendix-I}
\label{sec:appendix_sum_rate_constr}
\begin{theorem}
\label{thm:sum-rate-constraint} Consider a vector $(R_1, R_2,
\dots, R_n)$ such that \beqno
\begin{split} \sum_{i \in S} R_i &\geq H(X_S |X_{S^c}), \text{~for all $S
\subset \{1, 2, \dots , n\}$, and}\\ \sum_{i = 1}^n R_i &> H(X_1,
X_2, \dots, X_n). \end{split}\eeqno Then there exists another
vector $(R_1^{'}, R_2^{'}, \dots, R_n^{'})$ such that $R_i^{'}
\leq R_i$ for all $i = 1, 2, \dots n$ and \beqno
\begin{split} \sum_{i \in S} R_i^{'} &\geq H(X_S |X_{S^c}), \text{~for all $S
\subset \{1, 2, \dots , n\}$, and}\\ \sum_{i = 1}^n R_i^{'} &=
H(X_1, X_2, \dots, X_n). \end{split}\eeqno
\end{theorem}
\emph{Proof}. We claim that there exists a $R_{j^*} \in \{R_1,
R_2, \dots, R_n\}$ such that all inequalities in which $R_{j^*}$
participates are loose. 
The proof of this claim follows.
\par Suppose that the above claim is not true. Then for all $R_i$ where $i \in
\{ 1, 2, \dots, n\}$, there exists at least one subset $S_i
\subset \{1, 2, \dots, n\}$ such that, \beqno \sum_{k \in S_i} R_k
= H(X_{S_i} | S_i^c). \eeqno i.e. each $R_i$ participates in at
least one inequality that is tight.

Consider the subsets $S_1$ and $S_2$, i.e. the subsets for which
the inequalities are tight for $R_1$ and $R_2$ respectively. We
have by assumption, \beq
\begin{split}
&\sum_{k \in S_1} R_k + \sum_{k \in S_2} R_k = \sum_{k \in S_1
\cap S_2}
R_k + \sum_{k \in S_1 \cup S_2} R_k \\
&= H(X_{S_1} |X_{S_1^c}) + H(X_{S_2} |X_{S_2^c})\\
&\leq H(X_{S_1 \cup S_2} |X_{(S_1 \cup S_2)^c}) + H(X_{S_1 \cap
S_2} |X_{(S_1 \cap S_2)^c})
\end{split}
\eeq where in the second step we have used the supermodularity
property of conditional entropy. Now we are also given that \beq
\sum_{k \in S_1 \cap S_2} R_k \geq H(X_{(S_1 \cap S_2)} | X_{(S_1
\cap S_2)^c}). \eeq Therefore we can conclude that \beq \sum_{k
\in S_1 \cup S_2} R_k \leq H(X_{(S_1 \cup S_2)} | X_{(S_1 \cup
S_2)^c}).\eeq Now let $S_{12} = S_1 \cup S_2$. We have two cases
\begin{itemize}
\item[a)] $S_{12} = \{1, 2, \dots, n \}$.\\
In this case we have a contradiction since the conclusion above
implies $\sum_{i = 1}^n R_i \leq H(X_1, X_2, \dots, X_n)$.
\item[b)] $S_{12} \subset \{1, 2, \dots, n \}$.\\
In this case consider applying a similar argument as before with
$S_{12}$ and $S_3$. i.e. \beq
\begin{split}
&\sum_{k \in S_{12}} R_k + \sum_{k \in S_3} R_k = \sum_{S_{12}
\cup S_3} R_k + \sum_{S_{12} \cap S_3} R_k\\
&\leq H(X_{(S_1 \cup S_2)} | X_{(S_1 \cup S_2)^c}) + H(X_{S_3} |
X_{S_3^c})\\
& \leq H(X_{(S_1 \cup S_2 \cup S_3)} | X_{(S_1 \cup S_2 \cup
S_3)^c}) \\
&+ H(X_{S_1 \cup S_2 \cap S_3} | X_{(S_1 \cup S_2 \cap S_3)^c}).
\end{split}
\eeq Now since \beqno \sum_{k \in S_{12} \cap S_3} R_k \geq
H(X_{S_1 \cup S_2 \cap S_3} | X_{(S_1 \cup S_2 \cap S_3)^c})
\eeqno
 we obtain \beqno \sum_{S_{1} \cup S_2 \cup S_3} R_k \leq
H(X_{(S_1 \cup S_2 \cup S_3)} | X_{(S_1 \cup S_2 \cup S_3)^c}).
\eeqno
\end{itemize}
If $S_1 \cup S_2 \cup S_3 = \{1, 2, \dots, n\}$ we have the
required contradiction otherwise we can we can argue recursively
to arrive at the contradiction. Note that the process terminates
since $S_1 \cup S_2 \dots \cup S_n = \{1, 2, \dots, n\}$.

\par The above argument shows that there exists some $j^*$ such
that all inequalities in which $R_{j^*}$ participates are loose.
Therefore we can reduce $R_{j^*}$ to a new value $R^{red}_{j^*}$
until one of the inequalities in which it participates is tight.
If the sum-rate constraint is met with equality then we can set
$R_{j^*}^{'} = R^{red}_{j^*}$ otherwise we can recursively apply
the theorem to arrive at a new vector that is component-wise
smaller that the original vector $(R_1, R_2, \dots, R_n)$.
\endproof

We refer the S-W constraint $\sum_{j} R_j \geq H(X_1, \dots,
X_{N_S})$ as the sum rate constraint. From theorem
\ref{thm:sum-rate-constraint} we realize that if there exists a
rate vector $(R_1, R_2, \dots, R_n) \in \mathcal{R_{SW}}$ that
does not meet the sum rate constraint with equality then we can
always find another vector $(R_1^{'}, R_2^{'}, \dots, R_n^{'})$
that is component-wise smaller and meets the sum-rate constraint
with equality. Now consider the constraint
(\ref{eq:min-cost-sw-flow-balance}) in the {\it
MIN-COST-SW-NETWORK} problem. Instead of setting the flow-balance
at $s^{*}$ to $H(X_1, \dots, X_{N_S})$ and at $T_k$ to $-H(X_1,
\dots, X_{N_S})$ we could have introduced the constraints
\begin{align}
\sum_{\{j|(i,j) \in E^{*}\}} x_{ij}^{(T_k)} - \sum_{\{j | (j,i)
\in E^{*}\}} x_{ji}^{(T_k)} &\geq \sum_{j} R_j^{T_k}, i = s^{*} \text{~and}, \nonumber\\
\sum_{\{j|(i,j) \in E^{*}\}} x_{ij}^{(T_k)} - \sum_{\{j | (j,i)
\in E^{*}\}} x_{ji}^{(T_k)} &\leq -\sum_{j} R_j^{T_k}, i = T_k
\nonumber
\end{align}
and attempted to the solve the resulting linear program (without
forcing the sum rate constraint to be satisfied with equality).
Suppose that this optimization has a feasible point $(z, x, R)$
where the rate allocation for some terminal does not satisfy the
sum rate constraint. Then based on the previous observation we can
conclude that we can replace $R$ by a new set of rate allocations
$R^{'}$ such that $R \succeq R^{'}$ and $(z, x, R^{'})$ continues
to be feasible with the same cost. In fact one can possibly find
a new set of flows that may have a lower cost. To conclude this
shows that it is sufficient to consider rate allocations that
satisfy the sum rate constraint.

\section{Appendix-II}
\label{sec:appendix_lagrangian_deriv}
Consider the quadratic Gaussian CEO problem
with $n$ sources. For a given $(r_1, \dots, r_n)$, it can be seen
that the rank function of the contra-polymatroid specified by
$\mathcal{R(D)}$ is given by
\begin{equation*}
f(A) = \sum_{k \in A} r_k + \frac{1}{2} \log
\frac{\frac{1}{\sigma^2_X} + \sum_{k = 1}^{n} \frac{1 -
e^{-2r_k}}{\sigma^2_k}}{ \frac{1}{\sigma^2_X} + \sum_{k \in A^c}
\frac{1 - e^{-2r_k}}{\sigma^2_k}}
\end{equation*}
where $A \subseteq \{1, 2, \dots, n\}, A \neq \phi$. Therefore the
minimizer of (\ref{opt-contra-ceo}) can be written as
\begin{equation*}
\begin{split}
R_{\pi(1)} &= f(\{\pi(1)\}) \nonumber\\
R_{\pi(i)} &= f(\{\pi(1), \dots, \pi(i)\}) - f(\{\pi(1), \dots,
\pi(i-1)\}) \nonumber\\
\text{for~} &2\leq i \leq n \nonumber
\end{split}
\end{equation*}
where $\pi$ is a permutation such that $w_{\pi(1)} \geq
w_{\pi(2)}\geq \dots \geq w_{\pi(n)}$. The optimization problem in
(\ref{opt-contra-ceo}) becomes
\begin{equation}
\label{equiv-decomp-CEO}
\begin{split}
&\text{minimize~~} \sum_{i = 1}^{n} w_{\pi(i)} R_{\pi(i)}\\
&\text{subject to~~} \frac{1}{\sigma^2_X} + \sum_{i=1}^{N_S}
\frac{1 - e^{-2r_i}}{\sigma^2_i} \geq \frac{1}{D}, ~r_i \geq 0,
\forall i
\end{split}
\end{equation}
Let $A^{\pi}_i$ denote the set $\{\pi(1), \dots, \pi(i)\}^c$. The
objective function simplifies to
\begin{equation}
\label{obj_func_1} \sum_{i=1}^n w_{\pi(i)}r_{\pi(i)} + \frac{1}{2}
\sum_{i=1}^n w_{\pi(i)} \log \frac{\frac{1}{\sigma^2_X} + \sum_{k
\in A^{\pi}_{i-1}} \frac{1 - e^{-2r_k}}{\sigma^2_k}}{
\frac{1}{\sigma^2_X} + \sum_{k \in A^{\pi}_i} \frac{1 -
e^{-2r_k}}{\sigma^2_k}}
\end{equation}
We note that the first constraint in (\ref{equiv-decomp-CEO}) has
to be tight. To see this suppose $r_{\pi(1)} > 0$ and that the
constraint is not tight. Then we can reduce $r_{\pi(1)}$ so that
the objective function (\ref{obj_func_1}) is reduced. If
$r_{\pi(1)} = 0$, then the argument can be applied to $r_{\pi(2)}$
by realizing that the term corresponding to $i=1$ in the second
summation above is zero and so on. Using this observation the
optimization in (\ref{equiv-decomp-CEO}) can be rewritten as
\begin{equation}
\label{convex-decomp-CEO}
\begin{split}
&\text{minimize ~} \sum_{i=1}^n w_{\pi(i)}r_{\pi(i)} + w_{\pi(1)}
\log (1/D) - w_{\pi(n)}
\log(1/\sigma^2_X)\\
 &+ \sum_{i=0}^{n-1} (w_{\pi (i+1)} - w_{\pi
(i)}) \log \bigg{(} \frac{1}{\sigma^2_X} + \sum_{k \in A^{\pi}_i}
\frac{1 - e^{-2r_k}}{\sigma^2_k}\bigg{)}
\end{split}
\end{equation}
\vspace{-0.2cm}
\begin{equation}
\nonumber
\begin{split}
\text{subject to ~~} \frac{1}{D} - \frac{1}{\sigma^2_X} - \sum_{i
= 1}^n \frac{1 - e^{-2r_i}}{\sigma^2_i} &=
 0, \\
-r_i &\leq 0, \forall i.
\end{split}
\end{equation}

Next, we form the Lagrangian of the optimization problem in
(\ref{convex-decomp-CEO}) above with respect to the equality
constraint while treating the positivity constraint on the $r_i$'s
to be implicit and obtain the KKT conditions \cite{boyd_van}.
\beqno
\begin{split}
L(r, \nu) &=  \sum_{i=1}^n w_{\pi(i)}r_{\pi(i)} + w_{\pi(1)} \log
(1/D) - w_{\pi(n)}
\log(1/\sigma^2_X)\\
 &+ \sum_{i=1}^{n-1} (w_{\pi (i+1)} - w_{\pi
(i)}) \log \bigg{(} \frac{1}{\sigma^2_X} + \sum_{k \in A^{\pi}_i}
\frac{1 - e^{-2r_k}}{\sigma^2_k}\bigg{)} \\ &- \nu
\bigg{(}\frac{1}{\sigma^2_X} + \sum_{i = 1}^n \frac{1 -
e^{-2r_i}}{\sigma^2_i} - \frac{1}{D} \bigg{)}
\end{split}
\eeqno Differentiating with respect to $r_{\pi(i)}$ for $i = 1,
\dots, n$ and setting to zero, we obtain the following equations
\begin{eqnarray*}
\begin{split}
\frac{\partial L}{\partial r_{\pi(1)}} &= w_{\pi(1)} - \nu
\frac{2}{\sigma^2_{\pi(1)}} e^{-2 r_{\pi(1)}} = 0\\
\frac{\partial L}{\partial r_{\pi(k)}} &= w_{\pi(k)} - \frac{2
\nu}{\sigma^2_{\pi(k)}} e^{-2 r_{\pi(k)}} \\ &+ \frac{e^{-2
r_{\pi(k)}}}{\sigma^2_{\pi(k)}} \sum_{i=1}^{k-1}
\frac{w_{\pi(i+1)} - w_{\pi(i)}}{\frac{1}{D} - \sum_{\{\pi(1),
\dots , \pi(i)\}} \frac{1 - e^{-2 r_j}}{\sigma^2_j}} = 0
\end{split}
\end{eqnarray*}
Solving these equations, we obtain
\begin{eqnarray*}
\begin{split}
&r_{\pi(1)} = \bigg{[}\frac{1}{2} \log \bigg{(}
\frac{2\nu}{w_{\pi(1)} \sigma^2_{\pi(1)}}\bigg{)} \bigg{]}^{+}\\
&r_{\pi(k)} = \bigg{[}\frac{1}{2} \log \bigg{(}
\frac{2\nu}{w_{\pi(k)} \sigma^2_{\pi(k)}} ~-\\
&\frac{\sum_{i=1}^{k-1} (w_{\pi(i+1)} -
w_{\pi(i)})\big{[}\frac{1}{D} - \sum_{j \in \{\pi(1), \dots,
\pi(i)\}} \frac{1 - e^{-2 r_j}}{\sigma^2_j}\big{]}^{-1}
}{w_{\pi(k)} \sigma^2_{\pi(k)}}\bigg{)} \bigg{]}^{+} \\&\text{~
for $k \geq 2$} \end{split}
\end{eqnarray*}
where $x^{+} =\max(x, 0)$. Note that the form of the equations is
such that they can be solved recursively for a given value of
$\nu$. Furthermore $\nu$ is uniquely determined by the equality
constraint. Therefore a simple grid search on $\nu$ suffices to
solve these equations quickly. We note that a derivation similar
to the one above has been performed in a completely different
context in \cite{chenB08}.


\end{document}